\documentclass[a4paper,12pt]{article}
\usepackage{epsfig}
\usepackage[dvips,usenames]{color}
\usepackage{axodraw}
\usepackage{graphicx}

\newlength{\dinwidth}
\newlength{\dinmargin}
\begin{document}
\title{$B\to \pi \rho$, $\pi \omega$ Decays in Perturbative QCD Approach}
\author{Cai-Dian L\"u,
Mao-Zhi Yang\\
{\small CCAST (World Laboratory), P.O. Box 8730, Beijing 100080, China;}\\
{\small Institute of High Energy Physics, CAS,
 P.O. Box 918(4), Beijing 100039, China}\footnote{Mailing address}\\
and\\
{\small Physics Department, Hiroshima University, Higashi-Hiroshima 739-8526,
Japan}  }

\maketitle
\begin{picture}(0,0)
       \put(375,320){HUPD-0012}
\end{picture}

\begin{abstract}
We calculate the branching ratios and CP asymmetries for $B^0 \to \pi^+\rho^-$,
$B^0 \to \rho^+\pi^-$, $B^+ \to \rho^+\pi^0$,
$B^+ \to \pi^+\rho^0$, $B^0 \to \pi^0\rho^0$, $B^+ \to \pi^+\omega$ and
$B^0 \to \pi^0\omega$
 decays, in a
perturbative QCD approach. In this approach, we calculate non-factorizable
and annihilation type contributions, in addition to the usual factorizable
contributions. Our result is in agreement with the measured branching ratio
of $B^0/\bar B^0 \to \pi^\pm\rho^\mp$, $B^\pm \to \pi^\pm\rho^0$, $\pi^\pm \omega$
 by CLEO and BABAR collaboration.
We also predict large CP asymmetries in these decays.
These channels are useful to determine the CKM angle $\phi_2$.
\end{abstract}

\bigskip


\newpage

\section{Introduction}

The rare decays of B mesons are getting more and more interesting,
since they are useful for search of CP violation and sensitive to new
physics.
The recent measurement of  $B\to \pi \rho$ and $\pi \omega$ decays by CLEO
Collaboration \cite{cleo} arouse more discussions on these decays
\cite{recent}.
The $B\to \pi\rho$, $\pi\omega$
decays which are helpful to the determination of
Cabbibo-Kobayashi-Maskawa (CKM)
unitarity triangle $\phi_2$ have been studied in the factorization approach
in detail \cite{akl1,cheng}.
In this paper, we would like to study the $B\to \pi\rho$ and
$\pi \omega$ decays in the
 perturbative QCD approach (PQCD), where we can calculate the
 non-factorizable contributions as corrections to the usual factorization
approach.

In the $B\to \pi\rho$, $\pi \omega$
 decays, the $B$ meson is heavy, sitting at rest.
It decays into two light mesons with large momenta. Therefore the light
mesons are moving very fast in the rest frame of $B$ meson.
In this case, the short distance hard process dominates the decay amplitude.
The reasons can be ordered as: first, because there are not many resonances
near the energy region of $B$ mass, so it is reasonable to assume that
final state interaction is not important in two-body $B$ decays.
Second, With the final light
mesons moving very fast, there must be a hard gluon to kick
the light spectator quark (almost at rest)
in the B meson to form a fast moving pion or light vector meson.
So the dominant diagram in this theoretical picture is that one hard gluon
from the spectator quark connecting with the other quarks in the four
quark operator of the weak interaction.
There are also soft (soft and collinear) gluon exchanges between quarks.
Summing over those leading soft contributions gives a Sudakov form
factor, which suppresses the soft contribution to be dominant.
Therefore, it makes the PQCD reliable in calculating the non-leptonic
decays. With the Sudakov resummation, we can include the leading
double logarithms
for all loop diagrams, in association with the soft contribution.
Unlike the usual factorization approach, the hard part of the PQCD approach
consists of six quarks rather than four.
We thus call it six-quark operators or six-quark effective theory.
Applying the six-quark effective theory to B meson decays, we need meson
wave functions for the hadronization of quarks into mesons.
All the collinear dynamics are included in the meson wave functions.

 In this paper, we calculate the $B\to \pi$ and
$B\to \rho$ form factors, which are input parameters used in factorization
approach. The form factor calculations can give severe restrictions to the
input meson wave functions.
We also calculate the non-factorizable contributions and
 the annihilation type diagrams, which are difficult to calculate
in the factorization approach.
 We  found that this type of diagrams give
dominant contributions to strong phases.
The strong phase in this approach can also be calculated directly, without
ambiguity.
In the next section, we will briefly introduce our method of PQCD.
In section 3, we perform the perturbative calculations for all the channels.
And we give the numerical results and discussions in section 4.
Finally section 5 is a short summary.

\section{The Frame Work}

The three scale
PQCD factorization theorem has been developed for non-leptonic heavy
meson decays \cite{li}.
The
factorization formula is given by the typical expression,
\begin{equation}
C(t) \times H(x,t) \times \Phi (x) \times \exp\left[ -s(P,b)
-2 \int _{1/b}^t \frac{ d \bar\mu}{\bar \mu} \gamma_q (\alpha_s (\bar \mu))
\right],\label{factf}
\end{equation}
where $C(t)$ are the corresponding Wilson coefficients,
$\Phi (x)$ are the  meson wave functions.
And the quark anomalous dimension $\gamma_q=-\alpha_s /\pi$ describes
the evolution from scale $t$ to $1/b$.

Non-leptonic heavy meson decays involve three
scales: the W boson mass $m_W$, at which the matching condition of the
effective Hamiltonian are defined,
the typical scale $t$ of a hard subamplitude, which reflects
the dynamics of heavy quark decays,
and the factorization scale $1/b$,
with b the conjugate variable of parton transverse momenta.
The dynamics below $1/b$ scale is regarded as being completely
non-perturbative, and can be parameterized into meson wave functions.
Above the scale $1/b$, PQCD is reliable and radiative corrections
produce two types of large logarithms:
$ \ln (m_W/t)$ and $\ln(tb)$.
The former are summed by renormalization group equations to give the
leading logarithm evolution from $m_W$ to $t$ scale contained in the
Wilson coefficients $C(t)$.
While the latter are summed to give the evolution
from $t$ scale down to $1/b$, shown as the last factor in eq.(\ref{factf}).

There exist also double logarithms $\ln^2(Pb)$ from the overlap of collinear
and soft divergence, $P$ being the dominant light-cone component of a
meson momentum.
The resummation of these double logarithms leads to a Sudakov form factor
$\exp[-s(P,b)]$, which suppresses the long distance contributions in the
large $b$ region,
and vanishes as $b> 1/\Lambda_{QCD}$.
This factor improves the applicability of PQCD.
For the detailed derivation of the Sudakov form factors, see
ref.\cite{7,8}.
Since all logarithm corrections have been summed by renormalization group
equations, the above factorization formula does not depend on the
renormalization scale $\mu$.

With all the large logarithms resummed, the remaining finite contributions are
absorbed into a hard   sub-amplitude $H(x,t)$. The $H(x,t)$  are calculated
perturbatively involving the four quark operators together with the
spectator quark, connected by a hard gluon. When the end-point region
($x\to 0,1$) of wave function is important for the hard amplitude, the
corresponding large
double logarithms $\alpha_s ln^2x$ shall appear in the hard amplitude
$H(x,t)$, which should be resummed to give a jet function $S_t(x)$. This
technique is the so-called threshold resummation \cite{threshold}. The
threshold resummation form factor $S_t(x)$ vanishes as $x\to 0,1$, which
effectively suppresses the end-point behavior of the hard amplitude. This
suppression will become important when the meson wave function remains
constant at the end-point region. For example, the twist-3 wave function
$\phi_{\pi}^P$ and $\phi_{\pi}^t$ are such kinds of wave functions,
which can be found in the numerical section of this paper.
     The typical scale $t$ in the hard sub-amplitude  is around
     $\sqrt{\Lambda M_B}$. It is chosen as the maximum value of
     those scales appeared in the six quark action.
     This is to diminish the $\alpha_s^2$ corrections to the six
     quark amplitude. The expression of scale $t$ in different
     sub-amplitude will be derived in the next section and the formula is
     showed
     in the appendix.

\subsection{Wilson Coefficients}

First we begin with the weak effective Hamiltonian $H_{eff}$
for the $\Delta B=1$ transitions as
\begin{equation}
\label{heff}
{\cal H}_{eff}
= \frac{G_{F}} {\sqrt{2}} \, \left[ V_{ub} V_{ud}^*
\left (C_1 O_1^u + C_2 O_2^u \right)
-
V_{tb} V_{td}^* \,
\sum_{i=3}^{10}
C_{i} \, O_i   \right] \quad .
\end{equation}
We
specify below the operators in  ${\cal H}_{eff}$ for $b \to d$:
\begin{equation}\begin{array}{llllll}
 O_1^{u} & = &  \bar d_\alpha\gamma^\mu L u_\beta\cdot \bar
u_\beta\gamma_\mu L b_\alpha\ ,
&O_2^{u} & = &\bar d_\alpha\gamma^\mu L u_\alpha\cdot \bar
u_\beta\gamma_\mu L b_\beta\ , \\
O_3 & = & \bar d_\alpha\gamma^\mu L b_\alpha\cdot \sum_{q'}\bar
 q_\beta'\gamma_\mu L q_\beta'\ ,   &
O_4 & = & \bar d_\alpha\gamma^\mu L b_\beta\cdot \sum_{q'}\bar
q_\beta'\gamma_\mu L q_\alpha'\ , \\
O_5 & = & \bar d_\alpha\gamma^\mu L b_\alpha\cdot \sum_{q'}\bar
q_\beta'\gamma_\mu R q_\beta'\ ,   &
O_6 & = & \bar d_\alpha\gamma^\mu L b_\beta\cdot \sum_{q'}\bar
q_\beta'\gamma_\mu R q_\alpha'\ , \\
O_7 & = & \frac{3}{2}\bar d_\alpha\gamma^\mu L b_\alpha\cdot
\sum_{q'}e_{q'}\bar q_\beta'\gamma_\mu R q_\beta'\ ,   &
O_8 & = & \frac{3}{2}\bar d_\alpha\gamma^\mu L b_\beta\cdot
\sum_{q'}e_{q'}\bar q_\beta'\gamma_\mu R q_\alpha'\ , \\
O_9 & = & \frac{3}{2}\bar d_\alpha\gamma^\mu L b_\alpha\cdot
\sum_{q'}e_{q'}\bar q_\beta'\gamma_\mu L q_\beta'\ ,   &
O_{10} & = & \frac{3}{2}\bar d_\alpha\gamma^\mu L b_\beta\cdot
\sum_{q'}e_{q'}\bar q_\beta'\gamma_\mu L q_\alpha'\ .
\label{operators}
\end{array}
\end{equation}
Here $\alpha$ and $\beta$ are the $SU(3)$ color indices;
$L$ and $R$ are the left- and right-handed projection operators with
$L=(1 - \gamma_5)$, $R= (1 + \gamma_5)$.
The sum over $q'$ runs over the quark fields that are active at the scale
$\mu=O(m_b)$, i.e., $(q'\epsilon\{u,d,s,c,b\})$.

The PQCD approach works well for the leading twist approximation and leading
double logarithm summation. For the Wilson coefficients, we will also use
the leading logarithm summation for the QCD corrections, although the
next-to-leading order calculations already exist in the literature
\cite{buras}.
This is the consistent way to cancel the explicit $\mu$ dependence in the
theoretical formulae.

If the scale $ m_b< t< m_W$, then we evaluate the Wilson coefficients at $t$
scale using leading logarithm running equations \cite{buras}, in the appendix B
of ref.\cite{luy}.
In numerical calculations, we use
  $\alpha_s=
4\pi/[\beta_1 \ln(t^2/{\Lambda_{QCD}^{(5)}}^2)]$ which is the leading order
expression with $\Lambda_{QCD}^{(5)}=193$MeV, derived
from $\Lambda_{QCD}^{(4)}=250$MeV.
Here $\beta_1=(33-2n_f)/12$, with the
appropriate number of active quarks $n_f$.
$n_f=5$ when scale $t$ is larger than $m_b$.

If the scale $t < m_b$, then we evaluate the Wilson coefficients at $t$
scale using the  formulae in appendix C of ref.\cite{luy} for
four active quarks ($n_f=4$) (again in leading logarithm approximation).

\subsection{Wave Functions}

In the resummation procedures, the $B$ meson is treated as a heavy-light
system. In general, the B meson light-cone matrix element can be
decomposited as \cite{grozin,bene}
\begin{eqnarray}
&&\int_0^1\frac{d^4z}{(2\pi)^4}e^{i\bf{k_1}\cdot z}
   \langle 0|\bar{b}_\alpha(0)d_\beta(z)|B(p_B)\rangle \nonumber\\
&=&-\frac{i}{\sqrt{2N_c}}\left\{(\not p_B+m_B)\gamma_5
\left[\phi_B ({\bf k_1})-\frac{\not n-\not v}{\sqrt{2}}
\bar{\phi}_B({\bf k_1})\right]\right\}_{\beta\alpha},
\label{aa1}
\end{eqnarray}
where $n=(1,0,{\bf 0_T})$, and $v=(0,1,{\bf 0_T})$ are the unit vectors
pointing to the plus and minus directions, respectively. From the above
equation, one can see that there are two Lorentz structures in the B meson
distribution amplitudes. They obey to the following normalization
conditions
\begin{equation}
\int \frac{d^4 k_1}{(2\pi)^4}\phi_B({\bf k_1})=\frac{f_B}{2\sqrt{2N_c}},
~~~\int \frac{d^4 k_1}{(2\pi)^4}\bar{\phi}_B({\bf k_1})=0.
\end{equation}
In general, one should consider both these two Lorentz structures in
calculations of $B$ meson decays. However, it can be argued that the
contribution of $\bar{\phi}_B$ is numerically small \cite{kurimoto},
thus its contribution can be numerically neglected. Therefore, we only
consider the contribution of Lorentz structure
\begin{equation}
\Phi_B= \frac{1}{\sqrt{2N_c}} (\not p_B +m_B) \gamma_5 \phi_B ({\bf k_1}),
\label{bmeson}
\end{equation}
in our calculation. We keep the same input as other calculations
in this direction \cite{luy,kurimoto,keum}
and it is also easier for comparing with other approaches \cite{bene,bbns}.
Through out this paper, we use the light-cone coordinates to write the
four momentum as ($k_1^+,k_1^-, k_1^\perp$).
In the next section, we will see
that the hard part is always independent of one of the $k_1^+$ and/or $k_1^-$,
if we make some approximations.
The B meson wave function is then the function of  variable
$k_1^-$ (or $k_1^+$) and $k_1^\perp$.
\begin{equation}
\phi_B (k_1^-, k_1^\perp)=\int d k_1^+ \phi (k_1^+, k_1^-, k_1^\perp).
\label{int}
\end{equation}

The $\pi$ meson is treated as a light-light system.
At the $B$ meson rest frame,
pion is moving very fast,  one  of $k_1^+$ or
$k_1^-$ is zero depends on the definition of the z axis.
We consider a pion moving in the minus direction in this
paper. The pion distribution amplitude is defined by
\cite{ball}
\begin{eqnarray}
&&<\pi^-(P)|\bar{d_{\alpha}}(z)u_{\beta}(0)|0>\nonumber\\
&=&\frac{i}{\sqrt{2N_c}}\int_0^1
e^{ixP\cdot z}\left[\gamma_5\not
P\phi_{\pi}(x) +m_0\gamma_5\phi_P(x)
-m_0\sigma^{\mu\nu}\gamma_5 P_{\mu}z_{\nu}
\frac{\phi_{\sigma}(x)}{6}\right]_{\beta\alpha}
\label{aa2}
\end{eqnarray}
For the first and second terms in the above equation,
we can easily get the projector of the distribution
amplitude in the momentum space. However, for the
third term we should make some effort to transfer it
into the momentum space. By using integration by parts
for the third term, after a few steps, eq.(\ref{aa2})
can be finally changed to be
\begin{eqnarray}
&&<\pi^-(P)|\bar{d_{\alpha}}(z)u_{\beta}(0)|0>\nonumber\\
&=&\frac{i}{\sqrt{2N_c}}\int_0^1
e^{ixP\cdot z}\left[\gamma_5\not
P\phi_{\pi}(x) +m_0\gamma_5\phi_P(x)
+m_0[\gamma_5(\not v\not n-1)]\phi_\pi^t(x)\right]_{\beta\alpha}
\end{eqnarray}
where $\phi_\pi^t(x)=\frac{1}{6}\frac{d}{x}\phi_{\sigma}(x)$,
and vector $v$ is parallel to the $\pi$ meson momentum $p_\pi$.
 And $m_0=m_\pi^2/(m_u+m_d)$
 is a scale characterized by the Chiral
perturbation theory. In $B\to\pi\rho$ decays, $\rho$ meson is
only longitudinally polarized. We only consider its wave function
in longitudinal polarization \cite{kurimoto,ball2}.
\begin{equation}
<\rho^-(P,\epsilon_L)|\bar{d_{\alpha}}(z)u_{\beta}(0)|0>=
\frac{1}{\sqrt{2N_c}}\int_0^1e^{ixP\cdot z}
\left\{ \not \epsilon \left[\not p_\rho
\phi_\rho^t (x) + m_\rho \phi_\rho (x) \right]
+m_\rho \phi_\rho^s (x)\right\}
\end{equation}
The second term in the above equation is the leading twist wave function
(twist-2),
 while the first and third terms are sub-leading twist (twist-3) wave functions.

The transverse momentum $k^\perp$ is usually conveniently
converted to the $b$ parameter by
Fourier transformation.
 The initial conditions of $\phi_i(x)$,
$i=B$, $\pi$, are of non-perturbative origin, satisfying the
normalization
\begin{equation}
\int_0^1\phi_i(x,b=0)dx=\frac{1}{2\sqrt{6}}{f_i}\;,
\label{no}
\end{equation}
with $f_i$ the meson decay constants.

\section{Perturbative Calculations}

In the previous section
 we have discussed the wave functions and Wilson coefficients of
the factorization formula in eq.(\ref{factf}).
In this section, we will calculate the hard part $H(t)$.
This part involves the four quark operators and the necessary hard
gluon connecting the four quark operator and the spectator quark.
Since the final results are expressed as integrations of the distribution
function variables, we will show the whole amplitude for each diagram
including wave functions.

Similar to the $B\to \pi\pi$ decays \cite{luy},
there are  8 type diagrams contributing to the $B\to \pi\rho$ decays,
which are shown in Figure 1.
Let's first calculate the usual factorizable diagrams (a) and (b).
Operators $O_1$, $O_2$, $O_3$,
 $O_4$, $O_9$, and $O_{10}$ are $(V-A)(V-A)$ currents,  the sum of their
amplitudes is given as
\begin{eqnarray}
F_e&=& 8 \sqrt{2} \pi C_F G_F
f_\rho m_\rho m_B^2 (\epsilon \cdot p_\pi)
\int_{0}^{1}d x_{1}d x_{2}\,\int_{0}^{\infty} b_1d b_1 b_2d b_2\,
\phi_B(x_1,b_1)
\nonumber \\
& &  \left\{
\left[(1+x_2)\phi^A_\pi(x_2,b_2)
 + r_\pi  (1-2x_2) \left(\phi^P_\pi(x_2,b_2) +\phi^\sigma_\pi(x_2,b_2)
 \right)
 \right] \alpha_s (t_e^1) \right.
\nonumber \\
& & h_e(x_1,x_2,b_1,b_2)\exp[-S_{ab}(t_e^1)]
\nonumber \\
& & \left. + 2 r_\pi \phi^P_\pi(x_2,b_2) \alpha_s (t_e^2)
h_e(x_2,x_1,b_2,b_1)\exp[-S_{ab}(t_e^2)] \right \}\;,
\label{b}
\end{eqnarray}
where $ r_\pi = m_0 / m_B = m_\pi^2 / [ m_B (m_u+m_d)]$; $C_F=4/3$
is a color factor.
The function $h_e$, the scales $t_e^i$
 and the Sudakov factors $S_{ab}$ are displayed in the appendix.
In the above equation, we do not include the Wilson coefficients of
the corresponding operators, which are process dependent.
They will be shown later in this section for different decay
channels.
The diagrams Fig.1(a)(b) are also the diagrams for the
$B\to \pi$ form factor $F_1^{B\to \pi}$.
Therefore we can extract  $F_1^{B\to \pi}$ from eq.(\ref{b}).
\begin{equation}
F_1^{B\to \pi} (q^2=0)=\frac{F_e}{\sqrt{2}   G_F f_\rho m _\rho
 (\epsilon \cdot p_\pi) }.   \label{bpif}
\end{equation}
The operators $O_5$, $O_6$, $O_7$, and $O_8$ have a structure of
$(V-A) (V+A)$.
In some decay channels, some of these operators contribute to the
decay amplitude in a factorizable way.
Since only the vector part of the (V+A) current contribute to the
vector meson production,
\begin{equation}
\langle \pi |V-A|B\rangle \langle \rho |V+A | 0 \rangle =
\langle \pi |V-A|B\rangle \langle \rho |V-A | 0 \rangle  ,
\end{equation}
 the result of these operators is the same
as eq.(\ref{b}).
In some other cases, we need to do Fierz transformation  for these
operators to get right color structure for factorization to work.
In this case, we get (S-P)(S+P) operators from (V-A)(V+A) ones.
  Because neither scaler nor pseudo-scaler density gives
contribution to a
  vector meson production $\langle \rho|S+P| 0\rangle =0$, we get
\begin{eqnarray}
F_e^{P}&=&  0.
\end{eqnarray}

\begin{figure}[htbp]
 \scalebox{0.7}{
 {
 \fbox{
   \begin{picture}(140,120)(-30,0)
    \ArrowLine(30,60)(0,60)
    \ArrowLine(60,60)(30,60)
    \ArrowLine(90,60)(60,60)
    \ArrowLine(30,20)(90,20)
    \ArrowLine(0,20)(30,20)
    \Gluon(30,20)(30,60){5}{4} \Vertex(30,20){1.5} \Vertex(30,60){1.5}
    \Line(58,62)(62,58)
    \Line(58,58)(62,62)
    \ArrowLine(45,105)(60,65)
    \ArrowLine(60,65)(75,105)
    \put(-20,35){$B$}
    \put(100,35){$\pi$}
    \put(58,110){$\rho$}
    \put(45,48){\small{}}
    \put(65,66){\small{}}
    \put(0,65){\small{$\bar{b}$}}
    \put(45,0){(a)}
 \end{picture}
 }}}
 \scalebox{0.7}{
 {
 \fbox{
   \begin{picture}(140,120)(-30,0)
      \ArrowLine(30,60)(0,60)
      \ArrowLine(60,60)(30,60)
      \ArrowLine(90,60)(60,60)
      \ArrowLine(60,20)(90,20)
      \ArrowLine(0,20)(60,20)
      \Gluon(60,20)(60,60){5}{4} \Vertex(60,20){1.5} \Vertex(60,60){1.5}
      \Line(28,62)(32,58)
      \Line(28,58)(32,62)
      \ArrowLine(15,105)(30,65)
      \ArrowLine(30,65)(45,105)
      \put(-20,35){$B$}
      \put(100,35){$\pi$}
      \put(28,110){$\rho$}
      \put(15,48){\small{}}
      \put(35,66){\small{}}
      \put(0,65){\small{$\bar{b}$}}
      \put(35,0){(b)}
 \end{picture}
 }}}
 \scalebox{0.7}{
 {
 \fbox{
   \begin{picture}(170,110)(-20,0)
      \ArrowLine(47,60)(0,60)
      \ArrowLine(100,60)(53,60)
      \ArrowLine(20,20)(100,20)
      \ArrowLine(0,20)(20,20)
      \ArrowLine(37,80)(47,60)
      \ArrowLine(27,100)(37,80)
      \ArrowLine(53,60)(73,100)
      \Vertex(20,20){1.5} \Vertex(37,80){1.5}
      \GlueArc(150,20)(130,152,180){4}{9}
      \put(-20,35){$B$}
      \put(110,38){$\pi$}
      \put(48,105){$\rho$}
      \put(15,48){\small{}}
      \put(55,48){\small{}}
      \put(0,65){\small{$\bar{b}$}}
      \put(125,0){(c),(d)}
      \put(130,50){$+$}
   \end{picture}
   \begin{picture}(140,110)(-20,0)
      \ArrowLine(47,60)(0,60)
      \ArrowLine(100,60)(53,60)
      \ArrowLine(80,20)(100,20)
      \ArrowLine(0,20)(80,20)
      \ArrowLine(27,100)(47,60)
      \ArrowLine(63,80)(73,100)
      \ArrowLine(53,60)(63,80)
      \Vertex(80,20){1.5} \Vertex(63,80){1.5}
      \GlueArc(-50,20)(130,0,28){4}{9}
      \put(-20,35){$B$}
      \put(110,38){$\pi$}
      \put(48,105){$\rho$}
      \put(15,48){\small{}}
      \put(55,48){\small{}}
      \put(0,65){\small{$\bar{b}$}}
   \end{picture}
 }}}

\vspace{0.7cm}

 \scalebox{0.7}{
 {
 \fbox{
   \begin{picture}(135,130)(0,-20)
      \put(115,0){
      \rotatebox{90}{
        \ArrowLine(47,60)(0,60)
        \ArrowLine(100,60)(53,60)
        \ArrowLine(20,20)(100,20)
        \ArrowLine(0,20)(20,20)
        \ArrowLine(37,80)(47,60)
        \ArrowLine(27,100)(37,80)
        \ArrowLine(53,60)(73,100)
        \Vertex(20,20){1.5} \Vertex(37,80){1.5}
        \GlueArc(150,20)(130,152,180){4}{9}
      }}
      \put(0,45){$B$}
      \put(75,-10){$\pi$}
      \put(75,105){$\rho$}
      \put(60,38){\small{}}
      \put(60,55){\small{}}
      \put(20,75){\small{$\bar{b}$}}
      \put(115,45){$+$}
      \put(115,-20){(e),(f)}
   \end{picture}
   \begin{picture}(110,130)(0,-20)
      \put(115,0){
      \rotatebox{90}{
        \ArrowLine(47,60)(0,60)
        \ArrowLine(100,60)(53,60)
        \ArrowLine(80,20)(100,20)
        \ArrowLine(0,20)(80,20)
        \ArrowLine(27,100)(47,60)
        \ArrowLine(63,80)(73,100)
        \ArrowLine(53,60)(63,80)
        \Vertex(80,20){1.5} \Vertex(63,80){1.5}
        \GlueArc(-50,20)(130,0,28){4}{9}
      }}
      \put(0,45){$B$}
      \put(75,-10){$\pi$}
      \put(75,105){$\rho$}
      \put(60,37){\small{}}
      \put(60,55){\small{}}
      \put(20,75){\small{$\bar{b}$}}
   \end{picture}
 }}}
 \scalebox{0.7}{
 {
 \fbox{
   \begin{picture}(130,120)(0,-20)
      \put(45,0){
      \rotatebox{90}{
        \ArrowLine(30,60)(0,60)
        \ArrowLine(60,60)(30,60)
        \ArrowLine(90,60)(60,60)
        \ArrowLine(30,20)(90,20)
        \ArrowLine(0,20)(30,20)
        \Gluon(30,20)(30,60){5}{4} \Vertex(30,20){1.5} \Vertex(30,60){1.5}
        \Line(58,62)(62,58)
        \Line(58,58)(62,62)
        \ArrowLine(45,105)(60,65)
        \ArrowLine(60,65)(75,105)
        \put(65,66){\small{}}
      }}
      \put(0,55){$B$}
      \put(80,95){$\rho$}
      \put(80,-10){$\pi$}
      \put(15,78){\small{$\bar{b}$}}
      \put(65,55){\small{}}
      \put(115,45){$+$}
      \put(110,-20){(g),(h)}
   \end{picture}
   \begin{picture}(110,120)(0,-20)
      \put(45,0){
      \rotatebox{90}{
        \ArrowLine(30,60)(0,60)
        \ArrowLine(60,60)(30,60)
        \ArrowLine(90,60)(60,60)
        \ArrowLine(60,20)(90,20)
        \ArrowLine(0,20)(60,20)
        \Gluon(60,20)(60,60){5}{4} \Vertex(60,20){1.5} \Vertex(60,60){1.5}
        \Line(28,62)(32,58)
        \Line(28,58)(32,62)
        \ArrowLine(15,105)(30,65)
        \ArrowLine(30,65)(45,105)
        \put(35,66){\small{}}
      }}
      \put(0,25){$B$}
      \put(80,95){$\rho$}
      \put(80,-10){$\pi$}
      \put(15,48){\small{$\bar{b}$}}
      \put(65,25){\small{}}
   \end{picture}
 }}}
 \caption{Diagrams contributing to the $B\to \pi\rho$ decays (diagram
(a) and (b) contribute to the $B\to \pi$
form factor).}
\end{figure}
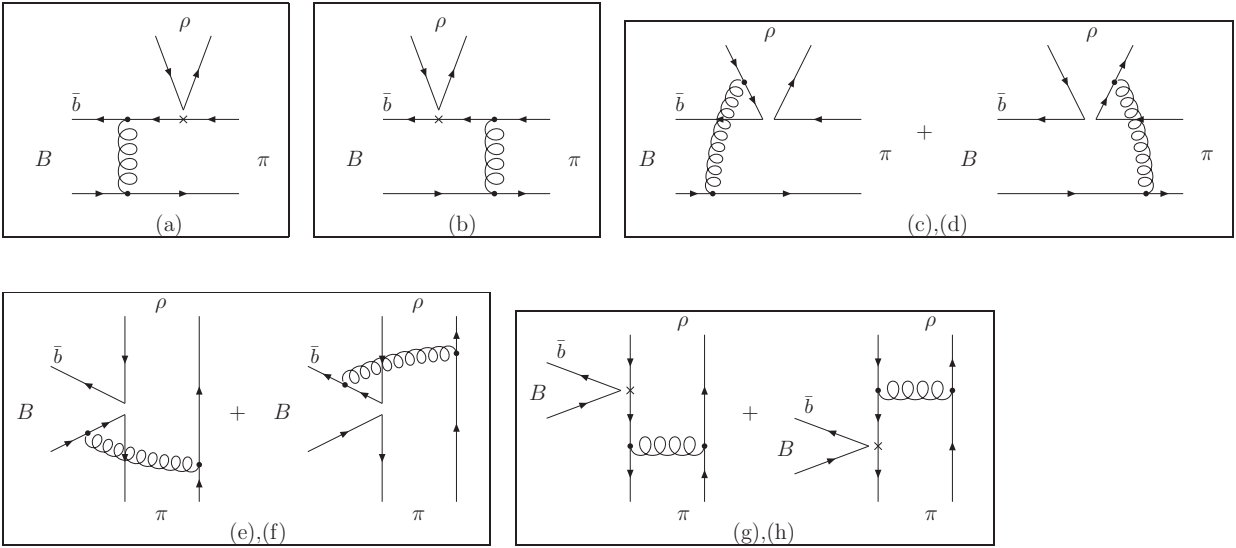

For the non-factorizable diagrams (c) and (d), all three meson wave functions
are involved.
The integration of $b_3$ can be performed easily using $\delta$ function
$\delta( b_3-b_1)$, leaving only integration of $b_1$ and $b_2$.
For the $(V-A)(V-A)$ operators the result is
\begin{eqnarray}
M_e&=& - \frac{32}{ 3}  \sqrt{3} \pi C_F G_F m_\rho m_B^2(\epsilon \cdot p_\pi)
\int_{0}^{1}d x_{1}d x_{2}\,d x_{3}\,\int_{0}^{\infty} b_1d b_1 b_2d b_2\,
\phi_B(x_1,b_1)
\nonumber \\
& & x_2~  \left[ \phi^A_\pi(x_2,b_1)-2r_\pi \phi^\sigma_\pi(x_2,b_1)\right]
 \phi_\rho(x_3,b_2)
 h_{d}(x_1,x_2,x_3,b_1,b_2)\exp[-S_{cd}(t_d)]\;.
\label{cd}
\end{eqnarray}
For the $(V-A)(V+A)$ operators the formula is different,
\begin{eqnarray}
M_e^P&=& \frac{64}{ 3} \sqrt{3} \pi C_F G_F m_\rho^2
 m_B (\epsilon \cdot p_\pi)
\int_{0}^{1}d x_{1}d x_{2}\,d x_{3}\,\int_{0}^{\infty} b_1d b_1 b_2d b_2\,
\phi_B(x_1,b_1)          \times
\nonumber \\
& & \left\{
 r_\pi (x_3-x_2) \left [  \phi^P_\pi(x_2,b_1)
\phi_\rho^t(x_3,b_2)+ \phi^\sigma_\pi(x_2,b_1) \phi_\rho^s(x_3,b_2)
\right]
\right.\nonumber\\
 &&  -r_\pi (x_2+x_3)\left[ \phi_\pi^P(x_2,b_1) \phi_\rho^s(x_3,b_2) +
   \phi_\pi^\sigma (x_2,b_1) \phi_\rho^t(x_3,b_2) \right]
\nonumber\\
    && \left. +x_3 \phi^A_\pi(x_2,b_1) \left[\phi_\rho^t(x_3,b_2)
     -
    \phi_\rho^s(x_3,b_2)\right]  \right\}
   h_{d}(x_1,x_2,x_3,b_1,b_2)\exp[-S_{cd}(t_d)]\;.
\label{cdp}
\end{eqnarray}
Comparing with the expression of $M_e$ in eq.(\ref{cd}), the
$(V-A)(V+A)$ type result $M_e^P$ is suppressed by $m_\rho/m_B$.

For the non-factorizable annihilation diagrams (e) and (f), again all three
wave functions are involved.
The integration of $b_3$ can be performed easily using $\delta$ function
$\delta( b_3-b_2)$.
Here we have two kind of contributions, which are different.
$M_a$ is contribution containing operator type $(V-A)(V-A)$, and
$M_a^P$ is contribution containing operator type $(V-A)(V+A)$.
\begin{eqnarray}
M_a &=& \frac{32}{ 3}  \sqrt{3} \pi C_F G_F m_\rho m_B^2(\epsilon \cdot p_\pi)
\int_{0}^{1}d x_{1}d x_{2}\,d x_{3}\,\int_{0}^{\infty} b_1d b_1 b_2d b_2\,
\phi_B(x_1,b_1)    \times
\nonumber \\
& & \left \{
 \left[ x_2\phi_\pi^A(x_2,b_2)\phi_\rho (x_3,b_2)  +r_\pi r_\rho (x_2-x_3)
 \left(\phi_\pi^P(x_2,b_2)  \phi_\rho^t (x_3,b_2)+
 \phi_\pi^\sigma(x_2,b_2)  \phi_\rho^s (x_3,b_2) \right)\right.
 \right.
 \nonumber\\
&&\left.+ r_\pi r_\rho (x_2+x_3)\left(\phi_\pi^\sigma(x_2,b_2)
 \phi_\rho^t (x_3,b_2)+
\phi_\pi^P(x_2,b_2)  \phi_\rho^s (x_3,b_2) \right)
\right] \nonumber\\
&&\times h_{f}^1(x_1,x_2,x_3,b_1,b_2)\exp[-S_{ef}(t_f^1)]
\nonumber  \\
&&
-\left[ x_3\phi_\pi^A(x_2,b_2)\phi_\rho (x_3,b_2)
+r_\pi r_\rho (x_3-x_2)\left( \phi_\pi^P(x_2,b_2)\phi_\rho^t (x_3,b_2)
+\phi_\pi^\sigma(x_2,b_2)\phi_\rho^s (x_3,b_2) \right)
\right.\nonumber\\
&&\left.
+r_\pi r_\rho (2+x_2+x_3) \phi_\pi^P(x_2,b_2)\phi_\rho^s (x_3,b_2)
-r_\pi r_\rho (2-x_2-x_3) \phi_\pi^\sigma(x_2,b_2)\phi_\rho^t (x_3,b_2)
\right]
\nonumber\\
&&
\left .  h_{f}^2(x_1,x_2,x_3,b_1,b_2)\exp[-S_{ef}(t_f^2)] \right \}
\;,    \label{e}
\\
M_a^P&=& -\frac{32}{ 3}  \sqrt{3} \pi C_F G_F m_\rho m_B^2
(\epsilon \cdot p_\pi)
\int_{0}^{1}d x_{1}d x_{2}\,d x_{3}\,\int_{0}^{\infty} b_1d b_1 b_2d b_2\,
\phi_B(x_1,b_1)  \times
\nonumber \\
& & \left \{
\left[  x_2 r_\pi  \phi_\rho (x_3,b_2) \left(\phi_\pi^P(x_2,b_2)
+\phi_\pi^\sigma(x_2,b_2)
\right)
 -x_3 r_\rho \phi_\pi^A(x_2,b_2)\left( \phi_\rho^t (x_3,b_2)+\phi_\rho^s (x_3,b_2)
 \right)
 \right]\right.\nonumber\\
 && \times  h_{f}^1(x_1,x_2,x_3,b_1,b_2)\exp[-S_{ef}(t_f^1)]
    \nonumber
\\
&&
+ \left[(2-x_2)r_\pi \phi_\rho (x_3,b_2)
\left(\phi_\pi^P(x_2,b_2)+\phi_\pi^\sigma(x_2,b_2)   \right)
   \right.      \label{ep}
 \\
&& \left . \left .-r_\rho (2-x_3)\phi_\pi^A(x_2,b_2)\left( \phi_\rho^t (x_3,b_2)
      +\phi_\rho^s (x_3,b_2) \right)    \right ]
 h_{f}^2(x_1,x_2,x_3,b_1,b_2)\exp[-S_{ef}(t_f^2)]
\right \} ,\nonumber
\end{eqnarray}
where $r_\rho= m_\rho/m_B$.
The factorizable annihilation diagrams (g) and (h) involve only
$\pi$ and $\rho$ wave functions.
There are also two kinds of decay amplitudes for these two diagrams.
$F_a$ is for $(V-A)(V-A)$ type operators, and
$F_a^P$ is for $(S-P)(S+P)$ type operators.
\begin{eqnarray}
F_a&=& 8  \sqrt{2} C_F G_F
\pi f_B m_\rho m_B^2(\epsilon \cdot p_\pi)
\int_{0}^{1}d x_{1}d x_{2}\,\int_{0}^{\infty} b_1d b_1 b_2d b_2
 \times  \nonumber \\
& & \left\{ \left[x_2\phi^A_\pi(x_1,b_1) \phi_\rho(x_2,b_2)
- 2 (1-x_2) r_\pi r_\rho
\phi^P_\pi(x_1,b_1)\phi_\rho^t(x_2,b_2)\right.\right.\nonumber\\
&&\left. +
       2 (1+x_2) r_\pi r_\rho
      \phi^P_\pi(x_1,b_1)\phi_\rho^s(x_2,b_2)
\right]  \alpha_s (t_e^1)
 h_{a}(x_2,x_1,b_2,b_1)
\exp[-S_{gh}(t_e^1)]
\nonumber \\
& &
-\left[x_1 \phi^A_\pi(x_1,b_1)  \phi_\rho(x_2,b_2)
  + 2 (1+x_1) r_\pi r_\rho
  \phi^P_\pi(x_1,b_1)\phi_\rho^s(x_2,b_2) \right. \nonumber\\
&& \left.\left . - 2 (1-x_1) r_\pi r_\rho
\phi^\sigma_\pi(x_1,b_1)\phi_\rho^s(x_2,b_2)  \right]
\alpha_s (t_e^2)
h_{a}(x_1,x_2,b_1,b_2)  \exp[-S_{gh}(t_e^2)]
\right \} \;,
\label{g}\\
F_a^P&=& 16  \sqrt{2} C_F G_F
\pi f_B m_\rho m_B^2 (\epsilon \cdot p_\pi)
\int_{0}^{1}d x_{1}d x_{2}\,\int_{0}^{\infty} b_1d b_1 b_2d b_2
\nonumber \\
& &\times \left\{ \left[2 r_\pi \phi^P_\pi(x_1,b_1)\phi_\rho(x_2,b_2)
 +x_2 r_\rho \phi^A_\pi(x_1,b_1) \left(\phi_\rho^s(x_2,b_2) -\phi_\rho^t(x_2,b_2)
 \right) \right
]\right .\nonumber \\
&&\alpha_s (t_e^1) h_{a}(x_2,x_1,b_2,b_1)
\exp[-S_{gh}(t_e^1)]\nonumber \\
& &
+\left[ x_1  r_\pi \left(\phi^P_\pi(x_1,b_1)
-   \phi^\sigma_\pi(x_1,b_1)\right)\phi_\rho(x_2,b_2)
+2r_\rho \phi_\pi^A(x_1,b_1)\phi_\rho^s(x_2,b_2) \right]  \nonumber \\
 &&\left. \alpha_s (t_e^2)
 h_{a}(x_1,x_2,b_1,b_2)  \exp[-S_{gh}(t_e^2)] \right \}\;,
\label{gp}
\end{eqnarray}

In the above equations, we have used the assumption that $x_1 <<x_2,x_3$.
Since the light quark momentum fraction $x_1$ in $B$ meson is peaked at
the small region, while quark momentum fraction $x_2$
 of
pion is peaked
around $0.5$, this is not a bad approximation.
The numerical results also show that this approximation makes very little
difference in the final result.
After using this approximation, all the diagrams are functions of
$k_1^-= x_1 m_B/\sqrt{2}$ of B meson only, independent of the variable of
$k_1^+$.
Therefore the integration of eq.(\ref{int}) is performed safely.

\begin{figure}[htbp]
 \scalebox{0.7}{
 {
 \fbox{
   \begin{picture}(140,120)(-30,0)
    \ArrowLine(30,60)(0,60)
    \ArrowLine(60,60)(30,60)
    \ArrowLine(90,60)(60,60)
    \ArrowLine(30,20)(90,20)
    \ArrowLine(0,20)(30,20)
    \Gluon(30,20)(30,60){5}{4} \Vertex(30,20){1.5} \Vertex(30,60){1.5}
    \Line(58,62)(62,58)
    \Line(58,58)(62,62)
    \ArrowLine(45,105)(60,65)
    \ArrowLine(60,65)(75,105)
    \put(-20,35){$B$}
    \put(100,35){$\rho$}
    \put(58,110){$\pi$}
    \put(45,48){\small{}}
    \put(65,66){\small{}}
    \put(0,65){\small{$\bar{b}$}}
    \put(45,0){(a)}
 \end{picture}
 }}}
 \scalebox{0.7}{
 {
 \fbox{
   \begin{picture}(140,120)(-30,0)
      \ArrowLine(30,60)(0,60)
      \ArrowLine(60,60)(30,60)
      \ArrowLine(90,60)(60,60)
      \ArrowLine(60,20)(90,20)
      \ArrowLine(0,20)(60,20)
      \Gluon(60,20)(60,60){5}{4} \Vertex(60,20){1.5} \Vertex(60,60){1.5}
      \Line(28,62)(32,58)
      \Line(28,58)(32,62)
      \ArrowLine(15,105)(30,65)
      \ArrowLine(30,65)(45,105)
      \put(-20,35){$B$}
      \put(100,35){$\rho$}
      \put(28,110){$\pi$}
      \put(15,48){\small{}}
      \put(35,66){\small{}}
      \put(0,65){\small{$\bar{b}$}}
      \put(35,0){(b)}
 \end{picture}
 }}}
 \scalebox{0.7}{
 {
 \fbox{
   \begin{picture}(170,110)(-20,0)
      \ArrowLine(47,60)(0,60)
      \ArrowLine(100,60)(53,60)
      \ArrowLine(20,20)(100,20)
      \ArrowLine(0,20)(20,20)
      \ArrowLine(37,80)(47,60)
      \ArrowLine(27,100)(37,80)
      \ArrowLine(53,60)(73,100)
      \Vertex(20,20){1.5} \Vertex(37,80){1.5}
      \GlueArc(150,20)(130,152,180){4}{9}
      \put(-20,35){$B$}
      \put(110,38){$\rho$}
      \put(48,105){$\pi$}
      \put(15,48){\small{}}
      \put(55,48){\small{}}
      \put(0,65){\small{$\bar{b}$}}
      \put(125,0){(c),(d)}
      \put(130,50){$+$}
   \end{picture}
   \begin{picture}(140,110)(-20,0)
      \ArrowLine(47,60)(0,60)
      \ArrowLine(100,60)(53,60)
      \ArrowLine(80,20)(100,20)
      \ArrowLine(0,20)(80,20)
      \ArrowLine(27,100)(47,60)
      \ArrowLine(63,80)(73,100)
      \ArrowLine(53,60)(63,80)
      \Vertex(80,20){1.5} \Vertex(63,80){1.5}
      \GlueArc(-50,20)(130,0,28){4}{9}
      \put(-20,35){$B$}
      \put(110,38){$\rho$}
      \put(48,105){$\pi$}
      \put(15,48){\small{}}
      \put(55,48){\small{}}
      \put(0,65){\small{$\bar{b}$}}
   \end{picture}
 }}}

\vspace{0.7cm}

 \scalebox{0.7}{
 {
 \fbox{
   \begin{picture}(135,130)(0,-20)
      \put(115,0){
      \rotatebox{90}{
        \ArrowLine(47,60)(0,60)
        \ArrowLine(100,60)(53,60)
        \ArrowLine(20,20)(100,20)
        \ArrowLine(0,20)(20,20)
        \ArrowLine(37,80)(47,60)
        \ArrowLine(27,100)(37,80)
        \ArrowLine(53,60)(73,100)
        \Vertex(20,20){1.5} \Vertex(37,80){1.5}
        \GlueArc(150,20)(130,152,180){4}{9}
      }}
      \put(0,45){$B$}
      \put(75,-10){$\rho$}
      \put(75,105){$\pi$}
      \put(60,38){\small{}}
      \put(60,55){\small{}}
      \put(20,75){\small{$\bar{b}$}}
      \put(115,45){$+$}
      \put(115,-20){(e),(f)}
   \end{picture}
   \begin{picture}(110,130)(0,-20)
      \put(115,0){
      \rotatebox{90}{
        \ArrowLine(47,60)(0,60)
        \ArrowLine(100,60)(53,60)
        \ArrowLine(80,20)(100,20)
        \ArrowLine(0,20)(80,20)
        \ArrowLine(27,100)(47,60)
        \ArrowLine(63,80)(73,100)
        \ArrowLine(53,60)(63,80)
        \Vertex(80,20){1.5} \Vertex(63,80){1.5}
        \GlueArc(-50,20)(130,0,28){4}{9}
      }}
      \put(0,45){$B$}
      \put(75,-10){$\rho$}
      \put(75,105){$\pi$}
      \put(60,37){\small{}}
      \put(60,55){\small{}}
      \put(20,75){\small{$\bar{b}$}}
   \end{picture}
 }}}
 \scalebox{0.7}{
 {
 \fbox{
   \begin{picture}(130,120)(0,-20)
      \put(45,0){
      \rotatebox{90}{
        \ArrowLine(30,60)(0,60)
        \ArrowLine(60,60)(30,60)
        \ArrowLine(90,60)(60,60)
        \ArrowLine(30,20)(90,20)
        \ArrowLine(0,20)(30,20)
        \Gluon(30,20)(30,60){5}{4} \Vertex(30,20){1.5} \Vertex(30,60){1.5}
        \Line(58,62)(62,58)
        \Line(58,58)(62,62)
        \ArrowLine(45,105)(60,65)
        \ArrowLine(60,65)(75,105)
        \put(65,66){\small{}}
      }}
      \put(0,55){$B$}
      \put(80,95){$\pi$}
      \put(80,-10){$\rho$}
      \put(15,78){\small{$\bar{b}$}}
      \put(65,55){\small{}}
      \put(115,45){$+$}
      \put(110,-20){(g),(h)}
   \end{picture}
   \begin{picture}(110,120)(0,-20)
      \put(45,0){
      \rotatebox{90}{
        \ArrowLine(30,60)(0,60)
        \ArrowLine(60,60)(30,60)
        \ArrowLine(90,60)(60,60)
        \ArrowLine(60,20)(90,20)
        \ArrowLine(0,20)(60,20)
        \Gluon(60,20)(60,60){5}{4} \Vertex(60,20){1.5} \Vertex(60,60){1.5}
        \Line(28,62)(32,58)
        \Line(28,58)(32,62)
        \ArrowLine(15,105)(30,65)
        \ArrowLine(30,65)(45,105)
        \put(35,66){\small{}}
      }}
      \put(0,25){$B$}
      \put(80,95){$\pi$}
      \put(80,-10){$\rho$}
      \put(15,48){\small{$\bar{b}$}}
      \put(65,25){\small{}}
   \end{picture}
 }}}
 \caption{Diagrams contributing to the $B\to \pi\rho$ decays (diagram
(a) and (b) contribute to the $B\to \rho$
form factor $A_0^{B\to \rho}$).}
\end{figure}
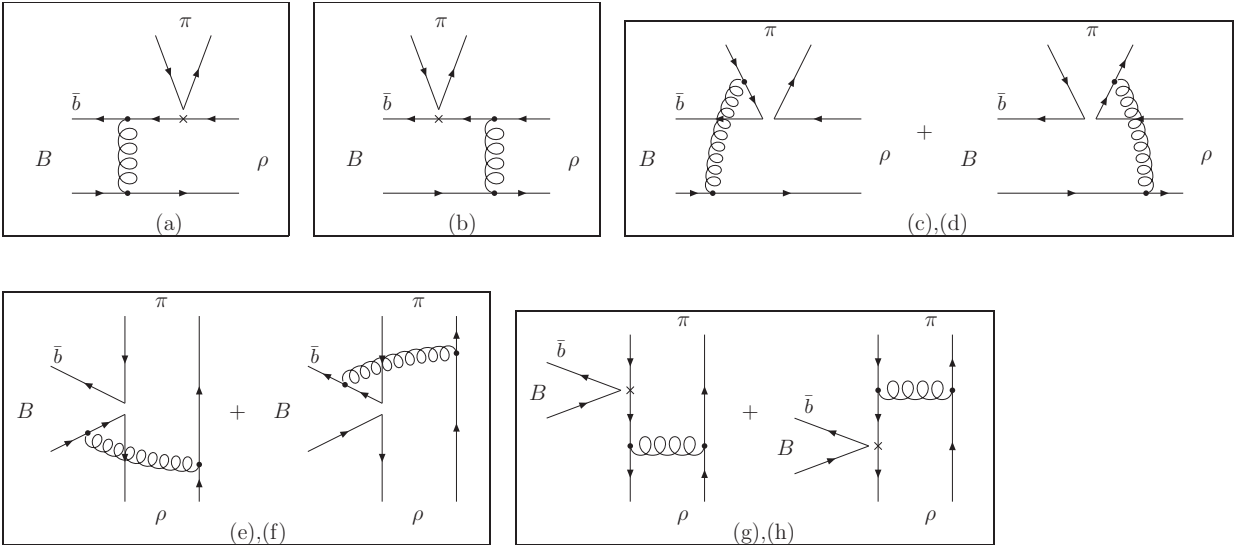

If we exchange the $\pi$ and $\rho$ in Figure 1, the result will be different
for some diagrams.
Because this will switch the dominant contribution
from $B\to \pi$ form factor to $B\to \rho$ form factor.
The new diagrams are shown in Figure 2.
Inserting $(V-A)(V-A)$ operators,
the corresponding amplitude for Figure 2(a)(b) is
\begin{eqnarray}
F_{e\rho}&=& 8 \sqrt{2} \pi C_F G_F
f_\pi m_\rho m_B^2 (\epsilon \cdot p_\pi)
\int_{0}^{1}d x_{1}d x_{2}\,\int_{0}^{\infty} b_1d b_1 b_2d b_2\,
\phi_B(x_1,b_1)
\nonumber \\
& &\times \left\{ \left[(1+x_2) \phi_\rho (x_2, b_2)
+(1-2x_2)r_\rho \left(\phi_\rho^t (x_2, b_2)+\phi_\rho^s (x_2, b_2)
\right)
\right] \right. \nonumber \\
&&\alpha_s(t_e^1) h_e(x_1,x_2,b_1,b_2)\exp[-S_{ab}(t_e^1)] \nonumber \\
&&\left.+2 r_\rho \phi_\rho^s (x_2, b_2)
\alpha_s(t_e^2) h_e(x_2,x_1,b_2,b_1)\exp[-S_{ab}(t_e^2)] \right\}
\;.
\label{aa}
\end{eqnarray}
These two diagrams are also responsible for the calculation of $B\to \rho$
form factors.
The form factor relative to the $B\to \pi \rho$ decays is $A_0^{B\to \rho}$,
which can be extracted from eq.(\ref{aa})
\begin{equation}
A_0^{B\to \rho} (q^2=0) = \frac{F_{e\rho}}{  \sqrt{2}    G_F
 f_\pi m_\rho  (\epsilon \cdot p_\pi) }.     \label{brhof}
\end{equation}
For $(V-A)(V+A)$ operators, Figure 2(a) and 2(b) give
\begin{eqnarray}
F_{e\rho}^{P}&=& -16 \sqrt{2} \pi C_F G_F
 f_\pi m_\rho  r_\pi m_B^2 (\epsilon \cdot p_\pi)
\int_{0}^{1}d x_{1}d x_{2}\,\int_{0}^{\infty} b_1d b_1 b_2d b_2\,
\phi_B(x_1,b_1)
\nonumber \\
& \times   &
 \left\{ \left[\phi_\rho (x_2, b_2)-r_\rho x_2 \phi_\rho^t (x_2, b_2)
 +(2+x_2) r_\rho \phi_\rho^s (x_2, b_2)\right]
 \right.
\nonumber \\
& &
  \times \alpha_s (t_e^1)  h_e (x_1,x_2,b_1,b_2)\exp[-S_{ab}(t_e^1)]
   \nonumber \\
   & &
 \left .+ \left[x_1 \phi_\rho (x_2, b_2) +2r_\rho \phi_\rho^s (x_2, b_2)\right]
\alpha_s (t_e^2)
 h_e(x_2,x_1,b_2,b_1)\exp[-S_{ab}(t_e^2)] \right\} \;,
\label{bb}
\end{eqnarray}
For the nonfactorizable diagrams Figure 2(c) and 2(d) the result is
\begin{eqnarray}
M_{e\rho}&=& -\frac{32}{ 3}
 \sqrt{3} \pi C_F G_F m_\rho m_B^2(\epsilon \cdot p_\pi)
\int_{0}^{1}d x_{1}d x_{2}\,d x_{3}\,\int_{0}^{\infty} b_1d b_1 b_2d b_2\,
\phi_B(x_1,b_1)
\nonumber \\
& &\times x_2\left[ \phi_\rho(x_2,b_2)-2r_\rho
\phi_\rho^t(x_2,b_2)\right]
\phi^A_\pi(x_3,b_1)
 h_{d}(x_1,x_2,x_3,b_1,b_2)\exp[-S_{cd}(t_d)]\;.
\label{cdcd}
\end{eqnarray}
For the nonfactorizable annihilation diagrams (e) and (f), we have
$M_{a\rho}$ for $(V-A)(V-A)$ operators and $M_{a\rho}^P$ for
$(V-A)(V+A)$ operators.
\begin{eqnarray}
M_{a\rho}&=& \frac{32}{ 3}
 \sqrt{3} \pi C_F G_F m_\rho m_B^2(\epsilon \cdot p_\pi)
\int_{0}^{1}d x_{1}d x_{2}\,d x_{3}\,\int_{0}^{\infty} b_1d b_1 b_2d b_2\,
\phi_B(x_1,b_1)       \left\{      \exp[-S_{ef}(t_f^1)] \right.
\nonumber \\
& &
\left[x_2 \phi_\pi^A(x_3,b_2) \phi_\rho(x_2,b_2)
+ r_\pi r_\rho (x_2-x_3) \left(\phi_\pi^P(x_3,b_2)  \phi_\rho^t(x_2,b_2)
+\phi_\pi^\sigma(x_3,b_2)  \phi_\rho^s(x_2,b_2) \right)
 \right.\nonumber \\
& &   \left.
  + r_\pi r_\rho (x_2+x_3) \left(\phi_\pi^\sigma(x_3,b_2)  \phi_\rho^t(x_2,b_2)
 + \phi_\pi^P(x_3,b_2)  \phi_\rho^s(x_2,b_2) \right)
 \right]
  h_f^1(x_1,x_2,x_3,b_1,b_2)   \nonumber \\
& &
-\left[ x_3 \phi_\pi^A(x_3,b_2)\phi_\rho(x_2,b_2) +r_\pi r_\rho (x_3-x_2)
\left( \phi_\pi^P(x_3,b_2)\phi_\rho^t(x_2,b_2)
+  \phi_\pi^\sigma(x_3,b_2)\phi_\rho^s(x_2,b_2)  \right) \right. \nonumber \\
& & \left .-r_\pi r_\rho (2-x_2-x_3) \phi_\pi^\sigma(x_3,b_2)
 \phi_\rho^t (x_2,b_2)
+r_\pi r_\rho (2+x_2+x_3) \phi_\pi^P(x_3,b_2) \phi_\rho^s(x_2,b_2) \right
]         \nonumber \\
&& \left . h_{f}^2(x_1,x_2,x_3,b_1,b_2) \exp[-S_{ef}(t_f^2)]] \right\}
\;,
\label{ee}\\
M_{a\rho}^P&=& M_a^P \;,
\label{eep}
\end{eqnarray}
For the factorizable annihilation diagrams (g) and (h)
\begin{eqnarray}
F_{a\rho}&=& -F_a,
\label{gg}
\\
F_{a\rho}^P&=& -F_a^P
\;,\label{ggp}
\end{eqnarray}
If the $\rho$ meson replaced by $\omega$ meson in Figure 1 and 2,
the formulas will be the same, except replacing $f_\rho$ by $f_\omega$
and $\phi_\rho$ replaced by $\phi_\omega$.

In the language of the above matrix elements for different
diagrams eq.(\ref{b}-\ref{ggp},),
the decay amplitude for $B^0\to \pi^+\rho^-$ can be written as
\begin{eqnarray}
{\cal M} (B^0 \to \pi^+\rho^-) &=& F_{e\rho} \left[ \xi_u \left(\frac{1}{3} C_1
+ C_2\right)-\xi_t
\left(C_4+\frac{1}{3}C_3 +C_{10}+\frac{1}{3}C_9\right)\right]\nonumber\\
&-& F_{e\rho}^P  \xi_t \left[
C_6+\frac{1}{3}C_5 +C_{8}+\frac{1}{3}C_7\right]\nonumber\\
&+& M_{e\rho} \left[\xi_u C_1-\xi_t (C_3+C_9)\right]\nonumber\\
&+& M_a  \left[\xi_u C_2- \xi_t \left( C_4 -C_6+\frac{1}{2}C_8
+ C_{10} \right)\right]
\nonumber\\
&-& M_{a\rho}  \xi_t \left[  C_3+C_4 -C_6-C_8
-\frac{1}{2}C_9- \frac{1}{2}C_{10} \right]
\nonumber\\
&-& M_{a\rho}^P \xi_t \left [ C_5- \frac{1}{2} C_7\right]
~+~ F_a \left[\xi_u \left(C_1+\frac{1}{3}C_2\right)\right.\nonumber
\\
&&\left.
-\xi_t\left( -\frac{1}{3}C_3-C_4
- \frac{3}{2}C_7 - \frac{1}{2} C_8 +\frac{5}{3}C_9 + C_{10}
\right) \right]\nonumber\\
&+& F_{a}^P \xi_t\left[  \frac{1}{3} C_5+ C_6
-\frac{1}{6} C_7 - \frac{1}{2} C_8 \right]
.\label{pm}
\end{eqnarray}
where $\xi_u = V_{ub}^*V_{ud}$, $\xi_t = V_{tb}^*V_{td}$.
The $C_i's$ should be calculated at the appropriate scale $t$ using
equations in the appendices of ref.\cite{luy}.
The decay amplitude of the charge conjugate decay channel
$\overline B^0\to \rho^+\pi^-$ is the same as eq.(\ref{pm}) except replacing
the CKM matrix elements $\xi_u$ to $\xi_u^*$ and $\xi_t$ to $\xi_t^*$
under the definition of charge conjugation $C|B^0\rangle = -
|\bar{B}^0 \rangle$.

\begin{eqnarray}
{\cal M} (B^0 \to \rho^+\pi^-) &=& F_e \left[ \xi_u \left(\frac{1}{3} C_1
+ C_2\right)-\xi_t
\left(C_4+\frac{1}{3}C_3 +C_{10}+\frac{1}{3}C_9\right)\right]\nonumber\\
&+& M_e \left[\xi_u C_1-\xi_t (C_3+C_9)\right]
-M_e^P \xi_t [C_5+C_7] \nonumber\\
&+& M_{a\rho}  \left[\xi_u C_2- \xi_t \left( C_4 -C_6+\frac{1}{2}C_8
+ C_{10} \right)\right]
\nonumber\\
&-& M_a \xi_t  \left[  C_3+C_4 -C_6- C_8
-\frac{1}{2}C_9- \frac{1}{2}C_{10} \right]
\nonumber\\
&-& M_a^P \xi_t \left[  C_5-\frac{1}{2}C_7\right]
~+~F_a \left[\xi_u \left(-C_1-\frac{1}{3}C_2\right)\right.
\nonumber\\
&& \left.-\xi_t\left( \frac{1}{3}C_3+C_4
+\frac{3}{2} C_7 + \frac{1}{2} C_8 -\frac{5}{3}C_9 -C_{10}
\right)\right]\nonumber\\
&-& F_a^P \xi_t\left[  \frac{1}{3} C_5+ C_6
-\frac{1}{6} C_7 - \frac{1}{2} C_8 \right].\label{pm2}
\end{eqnarray}

The decay amplitude for $B^0\to \pi^0\rho^0$ can be written as
\begin{eqnarray}
-{2}{\cal M} (B^0 \to \pi^0\rho^0) &=& F_e \left[
\xi_u \left(C_1+ \frac{1}{3}C_2\right) \right.\nonumber\\
&-& \left. \xi_t \left
(-\frac{1}{3}C_3 -C_4 +\frac{3}{2}C_7+\frac{1}{2}C_8+\frac{5}{3}C_9
+C_{10}\right)\right]\nonumber\\
&+& F_{e\rho} \left[
\xi_u \left(C_1+ \frac{1}{3}C_2\right) \right.\nonumber\\
&-& \left. \xi_t \left
(-\frac{1}{3}C_3 -C_4 -\frac{3}{2}C_7-\frac{1}{2}C_8+\frac{5}{3}C_9
+C_{10}\right)\right]
\nonumber\\
&+&  F_{e\rho}^P \xi_t\left[  \frac{1}{3} C_5+ C_6
-\frac{1}{6} C_7 - \frac{1}{2} C_8 \right]\nonumber\\
&+& M_e \left[\xi_u C_2-\xi_t (-C_3-\frac{3}{2}C_8+\frac{1}{2}C_9
+\frac{3}{2}C_{10}
)\right]\nonumber\\
&+& M_{e\rho} \left[\xi_u C_2-\xi_t (-C_3+\frac{3}{2}C_8+\frac{1}{2}C_9
+\frac{3}{2}C_{10}
)\right]\nonumber\\
&-& (M_a+M_{a\rho}) \left[\xi_u C_2\right .\nonumber\\
&&\left. - \xi_t\left( C_3 +2C_4 -2C_6
-\frac{1}{2}C_8 -\frac{1}{2}C_9+ \frac{1}{2}C_{10}\right) \right]
\nonumber\\
&+& (M_e^P+2M_a^P)  \xi_t\left[ C_5
-\frac{1}{2}C_7 \right].
\end{eqnarray}

The decay amplitude for $B^+\to \rho^+\pi^0$ can be written as
\begin{eqnarray}
\sqrt{2}{\cal M} (B^+ \to \rho^+\pi^0) &=& (F_e +2F_a)\left[
\xi_u \left(\frac{1}{3}C_1+ C_2\right)-\xi_t
\left(\frac{1}{3} C_3 +C_4+C_{10}+\frac{1}{3}C_9 \right)\right]\nonumber\\
&+& F_{e\rho} \left[
\xi_u \left(C_1+ \frac{1}{3}C_2 \right)-\xi_t
\left(-\frac{1}{3} C_3 -C_4-\frac{3}{2}C_7-\frac{1}{2}C_8
+C_{10}+\frac{5}{3}C_9 \right)\right]\nonumber\\
&-& F_{e\rho}^P  \xi_t \left[ -\frac{1}{3} C_5 - C_6+
\frac{1}{2}C_{8}+\frac{1}{6}C_7\right]\nonumber\\
&+& M_{e\rho} \left[\xi_u C_2-
\xi_t (-C_3+\frac{3}{2}C_8+\frac{1}{2}C_9+\frac{3}{2}C_{10}
)\right]
\nonumber\\
&+&(M_e+M_a-M_{a\rho}) \left[
\xi_u C_1 -\xi_t \left(C_3 +C_9 \right)\right]\nonumber\\
&-&M_e^P \xi_t \left[C_5 +C_7 \right]
~-~ 2F_a^P \xi_t\left[  \frac{1}{3} C_5+ C_6
+\frac{1}{3} C_7 + C_8 \right].\label{p0}
\end{eqnarray}

The decay amplitude for $B^+\to \pi^+\rho^0$ can be written as
\begin{eqnarray}
\sqrt{2}{\cal M} (B^+ \to \pi^+\rho^0) &=& F_e \left[
\xi_u \left(C_1+ \frac{1}{3}C_2\right)\right.\nonumber\\
&&~~~~\left.-\xi_t
\left(-\frac{1}{3} C_3-C_4 +\frac{3}{2}C_7 +\frac{1}{2}C_8
+\frac{5}{3}C_9+C_{10}\right)\right]\nonumber\\
&+& (F_{e\rho}-2F_a) \left[
\xi_u \left(\frac{1}{3}C_1+ C_2\right)-\xi_t
\left(\frac{1}{3} C_3+C_4
+\frac{1}{3}C_9+C_{10}\right)\right]\nonumber\\
&-& (F_{e\rho}^P -2F_a^P) \xi_t \left[ \frac{1}{3} C_5 +C_6+
\frac{1}{3}C_{7}+C_8\right]\nonumber\\
&+& M_e \left[\xi_u C_2-
\xi_t (-C_3-\frac{3}{2}C_8+\frac{1}{2}C_9+\frac{3}{2}C_{10})\right]\nonumber\\
&+& (M_{e\rho} -M_a +M_{a\rho})\left[\xi_u C_1-
\xi_t (C_3+C_9)\right]
\nonumber\\
&+& M_e^P \xi_t \left[C_5 -\frac{1}{2}C_7 \right].\label{p02}
\end{eqnarray}
 From eq.(\ref{pm}-\ref{p02}), we can verify that the isospin relation
\begin{equation}
{\cal M} (B^0 \to \pi^+\rho^-) +{\cal M} (B^0 \to \pi^-\rho^+)
-{2}{\cal M} (B^0 \to \pi^0\rho^0)
\end{equation}
$$= \sqrt{2}{\cal M} (B^+ \to \pi^0\rho^+)
+\sqrt{2}{\cal M} (B^+ \to \pi^+\rho^0) ,
$$
holds exactly in our calculations.

The decay amplitude for $B^+\to \pi^+\omega$ can also be written as
expressions of the above $F_i$ and $M_i$, but remember replacing $f_\rho$
by $f_\omega$ and $\phi_\rho$ by $\phi_\omega$.
\begin{eqnarray}
\sqrt{2}{\cal M} (B^+ \to \pi^+\omega) &=& F_e \left[
\xi_u \left(C_1+ \frac{1}{3}C_2\right)\right.\nonumber\\
&&~~~~\left.-\xi_t
\left(\frac{7}{3} C_3+\frac{5}{3} C_4 +2C_5+\frac{2}{3} C_6+
\frac{1}{2}C_7 +\frac{1}{6}C_8
+\frac{1}{3}C_9-\frac{1}{3} C_{10}\right)\right]\nonumber\\
&+& F_{e\rho} \left[
\xi_u \left(\frac{1}{3}C_1+ C_2\right)-\xi_t
\left(\frac{1}{3} C_3+C_4
+\frac{1}{3}C_9+C_{10}\right)\right]\nonumber\\
&-& F_{e\rho}^P  \xi_t \left[ \frac{1}{3} C_5 +C_6+
\frac{1}{3}C_{7}+C_8\right]\nonumber\\
&+& M_e \left[\xi_u C_2-
\xi_t (C_3+2C_4-2C_6-
\frac{1}{2}C_8-\frac{1}{2}C_9+\frac{1}{2}C_{10})\right]\nonumber\\
&+& (M_{e\rho} +M_a +M_{a\rho}) \left[
\xi_u C_1 -\xi_t \left(C_3 +C_9 \right)\right]\nonumber\\
&-&(M_a^P+M_{a\rho}^P) \xi_t \left[C_5 +C_7 \right]
~-~ M_e^P \xi_t \left[C_5-\frac{1}{2} C_7 \right]
.\label{omega}
\end{eqnarray}

The decay amplitude for $B^0\to \pi^0\omega$ can be written as
\begin{eqnarray}
{2}{\cal M} (B^0 \to \pi^0\omega) &=& F_e \left[
\xi_u \left(-C_1- \frac{1}{3}C_2\right) \right.\nonumber\\
&-& \left. \xi_t \left
(-\frac{7}{3}C_3 -\frac{5}{3}C_4 -2C_5-\frac{2}{3}C_6
-\frac{1}{2}C_7-\frac{1}{6}C_8-\frac{1}{3}C_9
+\frac{1}{3}C_{10}\right)\right]\nonumber\\
&+&F_{e\rho} \left[
\xi_u \left(C_1+ \frac{1}{3}C_2\right) \right.\nonumber\\
&-& \left. \xi_t \left
(-\frac{1}{3}C_3 -C_4 -\frac{3}{2}C_7-\frac{1}{2}C_8+\frac{5}{3}C_9
+C_{10}\right)\right]\nonumber\\
&+& F_{e\rho}^P  \xi_t \left[
C_6+\frac{1}{3}C_5 -\frac{1}{6}C_7-\frac{1}{2}C_{8}\right]\nonumber\\
&+& M_e \left[-\xi_u C_2-\xi_t (-C_3-{2}C_4+2C_6+\frac{1}{2}C_8
+\frac{1}{2}C_9-\frac{1}{2}C_{10}
)\right]\nonumber\\
&+& M_{e\rho} \left[\xi_u C_2-\xi_t (-C_3+\frac{3}{2}C_8
+\frac{1}{2}C_9+\frac{3}{2}
C_{10}
)\right]\nonumber\\
&+& (M_a+M_{a\rho}) \left[\xi_u C_2 - \xi_t\left( -C_3
-\frac{3}{2}C_8 +\frac{1}{2}C_9+ \frac{3}{2}C_{10}\right) \right]
\nonumber\\
&+& (M_e^P+2M_{a}^P)  \xi_t\left[ C_5
-\frac{1}{2}C_7 \right].\label{omega2}
\end{eqnarray}

\section{Numerical calculations and discussions of Results}

In the numerical calculations we use
$$
 \Lambda_{\overline{\mathrm{MS}}}^{(f=4)} = 0.25 { GeV}, f_\pi = 130
 { MeV}, f_B = 190 MeV,$$
$$
 m_0 = 1.4 { GeV}, f_\rho = f_\omega=200
 { MeV}, f_\rho^T =f_\omega^T = 160  MeV,$$
\begin{equation}
 M_B = 5.2792 { GeV}, M_W = 80.41{ GeV}.         \label{para}
\end{equation}
  Note, for simplicity, we use the same value for $f_\rho$
  ($f_\rho^T$) and $f_\omega$ ($f_\omega^T$).
  And this also makes it easy for us to see the major difference
  for the two mesons in the B decays.
  In principal, the decay constants can be a little different.
 For the light meson wave function, we neglect the $b$ dependence part, which is not
 important in numerical analysis.
 We use  wave function for $\phi_\pi^A$ and the twist-3 wave functions
 $\phi_{\pi}^P$ and $\phi_{\pi}^t$
  from \cite{ball}
 \begin{eqnarray}
 \phi_\pi^A(x) &=&  \frac{3}{\sqrt{6} }
  f_\pi  x (1-x)  \left[1+0.44C_2^{3/2} (2x-1) +0.25 C_4^{3/2}
  (2x-1)\right],\label{piw1}\\
 \phi_{\pi}^P(x) &=&   \frac{f_\pi}{2\sqrt{6} }
   \left[ 1+0.43 C_2^{1/2} (2x-1) +0.09 C_4^{1/2} (2x-1) \right]  ,\\
 \phi_{\pi}^t(x) &=&  \frac{f_\pi}{2\sqrt{6} } (1-2x)
   \left[ 1+0.55  (10x^2-10x+1)  \right]  .    \label{piw}
 \end{eqnarray}
 The Gegenbauer polynomials are defined by
 \begin{equation}
 \begin{array}{ll}
 C_2^{1/2} (t) = \frac{1}{2} (3t^2-1), & C_4^{1/2} (t) = \frac{1}{8}
 (35t^4-30t^2+3),\\
 C_2^{3/2} (t) = \frac{3}{2} (5t^2-1), & C_4^{3/2} (t) = \frac{15}{8}
 (21t^4-14t^2+1),
 \end{array}
 \end{equation}
whose coefficients correspond to $m_0=1.4$ GeV.
In $B\to \pi \rho $, $\pi\omega$ decays, it is the longitudinal polarization of
the $\rho$ and $\omega$ meson contribute to the decay amplitude.
Therefore we choose the wave function of $\rho$ and $\omega$ meson similar to the
pion case in eq.(\ref{piw1},\ref{piw}) \cite{ball2}
\begin{eqnarray}
\phi_\rho(x) ~=~ \phi_\omega(x)&=&  \frac{3}{\sqrt{6} }
 f_\rho  x (1-x)  \left[1+ 0.18C_2^{3/2} (2x-1) \right],\\
    \phi_\rho^t(x) ~=~ \phi_\omega^t(x)&=&  \frac{f_\rho^T }{2\sqrt{6} }
  \left\{  3 (2 x-1)^2 +0.3(2 x-1)^2  \left[5(2 x-1)^2-3  \right]
  \right.
  \nonumber\\
 &&~~\left. +0.21 [3- 30 (2 x-1)^2 +35 (2 x-1)^4] \right\},\\
\phi_\rho^s(x) ~=~ \phi_\omega^s(x)&=&  \frac{3}{2\sqrt{6} }
 f_\rho^T   (1-2x)  \left[1+ 0.76 (10 x^2 -10 x +1) \right] .
\end{eqnarray}
    Here again, for simplicity, we use the same expression for
    $\rho$ and $\omega$ mesons.

For $B$ meson, the wave function is chosen as
\begin{eqnarray}
\phi_B(x,b) &=&  \frac{N_B}{2\sqrt{6} } f_B x^2(1-x)^2 \mathrm{exp} \left
 [ -\frac{M_B^2\ x^2}{2 \omega_{b}^2} -\frac{1}{2} (\omega_{b} b)^2\right],
 \label{phib}
\end{eqnarray}
with  $\omega_{b}=0.4$ GeV.
$N_B=2365.57$ is a normalization factor.
We include full expression of twist-3 wave functions for light mesons unlike
$B\to \pi \pi$ decays \cite{luy}.
The twist-3 wave functions are also adopted from QCD sum rule
calculations \cite{ball}.
These changes make the $B\to \rho$ form factor a little larger than the
$B\to \pi$ form factor \cite{kurimoto}.
However, this set of parameters does not change
 the $B^0\to \pi^+ \pi^-$ branching ratios.
And we will see later that, this set of parameters will give good
results of $B\to \pi \rho$ and $\pi \omega$ decays. With the above
chosen wave functions, we get the corresponding form factors at
zero momentum transfer from eq.(\ref{bpif},\ref{brhof}):
$$F_0^{B\to \pi}
=0.30,~~~~~~A_0^{B\to \rho} =0.37.$$
They are close to the light
cone QCD sum rule results \cite{sum}.

The CKM parameters we used here are \cite{pdg}
\begin{equation}\begin{array}{ll}
|V_{ud}|=0.9735\pm 0.0008 &
|V_{ub}/V_{cb}|=0.090\pm 0.025 \\
|V_{cb}|=0.0405\pm 0.0019&
|V_{tb}^*V_{td}|=0.0083\pm 0.0016.
\end{array}
\end{equation}
We leave the CKM angle $\phi_2$ as a free parameter. $\phi_2$'s definition is
\begin{equation}
\phi_2=arg\left[-\frac{V_{td}V_{tb}^*}{V_{ud}V_{ub}^*}\right].
\end{equation}
In this parameterization, the decay amplitude of $B\to \pi\rho$ (or $\pi\omega$)
can be rewritten as
\begin{eqnarray}
{\cal M} &=& V_{ub}^*V_{ud} T -V_{tb}^* V_{td} P\nonumber\\
 &=& V_{ub}^*V_{ud} T
\left[1 +z e^{i(\phi_2+\delta)} \right],\label{m}
\end{eqnarray}
where $z=\left|\frac{V_{tb}^* V_{td}}{ V_{ub}^*V_{ud} } \right|
\left|\frac{P}{T} \right|$, and
$\delta$ is the relative strong phase between tree (T) diagrams and
penguin diagrams (P). $z$ and $\delta$ can be calculated from PQCD.
The corresponding charge conjugate decay mode is then
\begin{eqnarray}
\overline{\cal M} &=& V_{ub}V_{ud}^* T -V_{tb} V_{td}^* P\nonumber\\
 &=& V_{ub}V_{ud}^* T
\left[1 +z e^{i(-\phi_2+\delta)} \right].\label{mb}
\end{eqnarray}
Therefore the averaged branching ratio for $B\to \pi\rho$ is
\begin{eqnarray}
Br&=& (|{\cal M}|^2 +|\overline{\cal M}|^2)/2\nonumber\\
&=&  \left| V_{ub}V_{ud}^* T \right| ^2
\left[1 +2 z\cos \phi_2 \cos \delta +z^2 \right],\label{br}
\end{eqnarray}
where $z= \left|\frac{V_{tb} V_{td}^*}{ V_{ub}V_{ud}^* } \right|
\left|\frac{P}{T} \right|$.
 Eq.(\ref{br}) shows that the averaged branching ratio is a function
of $\cos \phi_2$.
This gives potential method to determine the CKM angle $\phi_2$ by measuring
only the averaged non-leptonic decay branching ratios.
In our PQCD approach, the branching ratios and also the other quantities
in eq.(\ref{br}) are all calculable, such that $\cos\phi_2$ is
measurable. However, there are still uncertainties in the input parameters
of our approach as discussions below. More experimental
data from BABAR and Belle  can restrict these parameters     in
the near future.
\begin{figure}\begin{center}
\epsfig{file=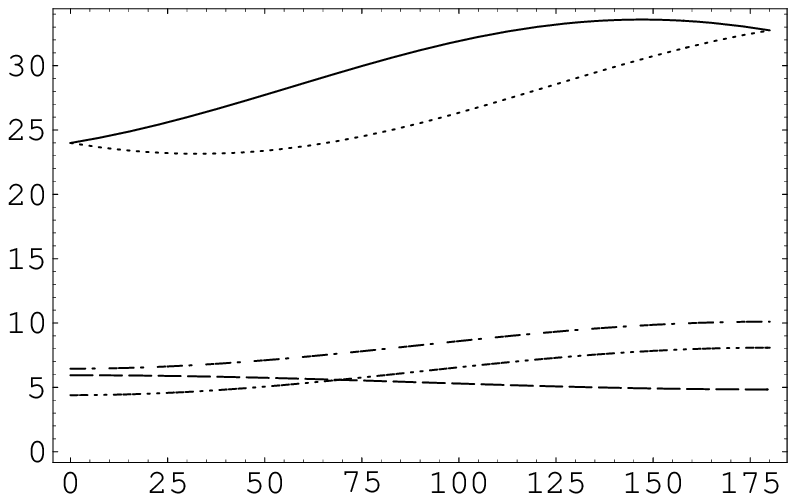,width=12cm}
\begin{picture}(0,0)(0,0)
   \put(-160,-10){$\phi_2$ (degree)}
   \put(-360,60){\rotatebox{90}{Br ($B\to \pi \rho$, $\pi\omega$) ($10^{-6}$)}}
  \end{picture}\end{center}
\caption{Branching ratios ($10^{-6}$)
 of $B^0/\bar B^0\to \pi^+ \rho^-$  (solid line),
 $B^0/\bar B^0\to \rho^+ \pi^-$ (dotted line), $B^+\to \pi^+ \rho^0$
(dashed line),
 $B^+\to \rho^+ \pi^0$ (dash-dotted line), and
 $B^+\to \pi^+ \omega$  (dash-dotted-dotted line),
as a function
of CKM angle $\phi_2$.}\label{pirho}
\end{figure}

More complicated, there are four decay channels of $B^0/\bar B^0\to \pi^+ \rho^-$,
$ B^0/\bar B^0\to \rho^+ \pi^-$.
Due to $B\bar B$ mixing, it is very difficult to distinguish
$B^0$ from $\bar B^0$. But it is very easy to identify the final states.
Therefore we sum up $B^0/\bar B^0\to \pi^+ \rho^-$ as one channel, and
$ B^0/\bar B^0\to \rho^+ \pi^-$ as another, although the summed up channels
are not charge conjugate states.
We show the branching ratio  of  $B^0/\bar B^0\to \pi^+ \rho^-$,
$ B^0/\bar B^0\to \rho^+ \pi^-$,
 $B^+\to \pi^+ \rho^0$,  $B^+\to \rho^+ \pi^0$, and
 $B^+\to \pi^+ \omega$  decays
as a function of $\phi_2$ in Figure~\ref{pirho}.
The branching ratio of $B^0/\bar B^0\to   \pi^+ \rho^-$ is a little
larger than that of $B^0/\bar B^0\to  \pi^-\rho^+$ decays.
Each of them is a sum of two decay channels. They are all  getting
 larger when $\phi_2$ is larger.
The average of the two is in agreement with
 the recently
  measured  branching ratios by CLEO \cite{cleo} and BABAR \cite{babar}
\begin{eqnarray}
 \mathrm{Br}({B}^0 \to \pi^+ \rho^-+\pi^- \rho^+) =
27.6 ^{+8.4}_{-7.4} \pm 4.2
 \times 10^{-6}, & CLEO\\
\mathrm{Br}({B}^0 \to \pi^+ \rho^-+\pi^- \rho^+) =
28.9 {\pm 5.4} \pm 4.3
 \times 10^{-6}, & BABAR
\end{eqnarray}
 There are still large uncertainties in the experimental results. Therefore
it is still early to fully determine the input parameters and to tell the
CKM angle $\phi_2$ from experiments.

    \begin{figure}\begin{center}
    \epsfig{file=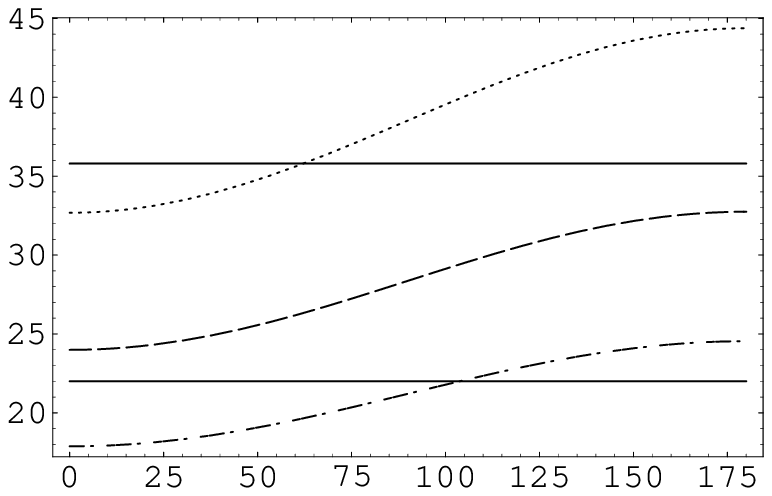,width=12cm}
    \begin{picture}(0,0)(0,0)
       \put(-160,-10){$\phi_2$ (degree)}
    \put(-360,60){\rotatebox{90}{Br ($B\to \pi \rho$) ($10^{-6}$)}}
      \end{picture}\end{center}
    \caption{Branching ratios of $B^0/\bar B^0\to
    \pi^\pm \rho^\mp$ decays: $\omega_b=0.36$
     (dotted line),    $\omega_b=0.40$
     (dashed line) and
    $\omega_b=0.44$ (dash-dotted line)
    as a function of CKM angle $\phi_2$. The two horizontal
    lines are BABAR measurements.}\label{om}
    \end{figure}
The most uncertain parameters in our approach are from the meson wave
functions. In principal, they can be only restricted by
experiments, namely, semi-leptonic   and non-leptonic decays of
B mesons. Our parameters chosen for the numerical calculations
in eq.(\ref{para},\ref{phib}) are  best fit values from
$B\to\pi\pi$ decays \cite{luy}, $B\to \pi K$ \cite{keum},
$B\to \pi$, $B\to \rho$ semi-leptonic decays \cite{kurimoto}
and some other experiments.
As in these decays, we show the $\omega_b$ dependence of
Branching ratios  $Br(B^0/\bar B^0\to\pi^\pm \rho^\mp)$
in Figure \ref{om}. The dotted, dashed and dash-dotted
lines are for $\omega_b=0.36$, $0.40$ and $0.44$, respectively.
They are also shown as a function of CKM angle $\phi_2$.
The two horizontal lines in the figure are
BABAR measurements of $1\sigma$.
 From the figure, we can see that the branching ratio is
quite sensitive to the $\omega_b$ parameter. Fortunately,
this parameter is also restricted from semi-leptonic decays
\cite{kurimoto}. In the near future, it will not be a big
problem for us.

\begin{figure}\begin{center}
    \epsfig{file=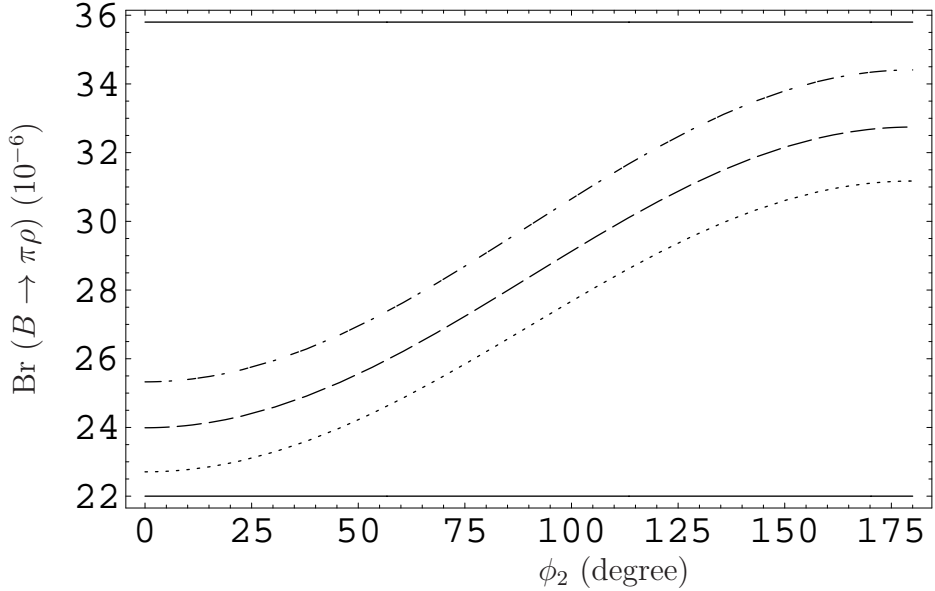,width=12cm}
    \begin{picture}(0,0)(0,0)
       \put(-160,-10){$\phi_2$ (degree)}
    \put(-360,60){\rotatebox{90}{Br ($B\to \pi \rho$) ($10^{-6}$)}}
      \end{picture}\end{center}
    \caption{Branching ratios of $B^0/\bar B^0\to
    \pi^\pm \rho^\mp$ decays: $m_0=1.3$ GeV
     (dotted line),    $m_0=1.4$   GeV
     (dashed line) and
    $m_0=1.5$ GeV (dash-dotted line)
    as a function
    of CKM angle $\phi_2$. The two horizontal
    lines are BABAR measurements.}\label{m0}
    \end{figure}

In Figure~\ref{m0}, we show the branching ratio of   $B^0/\bar B^0\to
\pi^\pm \rho^\mp$ decays: $m_0=1.3$ GeV
(dotted line), $m_0=1.4$ GeV (dashed line) and
$m_0=1.5$ GeV (dash-dotted line) as a function
of CKM angle $\phi_2$.
$m_0$ is a parameter characterize the relative size of twist 3
contribution from twist 2 contribution.
It originates from the Chiral perturbation theory and have a
value near 1 GeV. Because of the Chiral enhancement of $m_0$,
the twist 3 contribution is at the
same order magnitude as the twist 2 contribution. Thus
the branching ratios of $Br(B^0/\bar B^0\to
\pi^\pm \rho^\mp)$ are also
sensitive to this parameter,
 but not as strong as the $\omega_b$ dependence.

The branching ratios of
 $B^+\to \pi^+ \rho^0$ and $B^+\to \pi^+ \omega$ have little dependence
on $\phi_2$.
They are a little smaller than the CLEO measurement \cite{cleo}
showed below, but still within experimental error-bars.
\begin{equation}
 \mathrm{Br}({B}^+ \to \pi^+ \rho^0) = 10.4 ^{+3.3}_{-3.4} \pm 2.1
 \times 10^{-6},
\end{equation}
\begin{equation}
 \mathrm{Br}({B}^+ \to \pi^+ \omega) = 11.3 ^{+3.3}_{-2.9} \pm 1.4
 \times 10^{-6}.
\end{equation}
However, the recent BABAR measurement is in good agreement with
our prediction for    $B^+\to \pi^+ \omega$ \cite{piome}
\begin{equation}
 \mathrm{Br}({B}^+ \to \pi^+ \omega) = 6.6 ^{+2.1}_{-1.8} \pm 0.7
 \times 10^{-6},
\end{equation}
 where the error-bars are also smaller.
The preliminary result of Belle shows that the branching ratio of
${B}^+ \to \pi^+ \rho^0$ is around  $6\times 10^{-6}$ \cite{paoti}.
This  agrees with our prediction in Figure~\ref{pirho}.

\begin{figure}\begin{center}
\epsfig{file=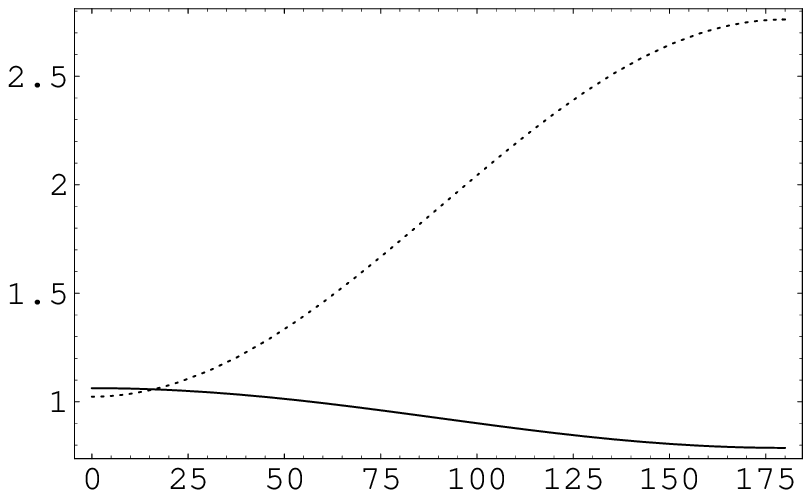,width=12cm}
\begin{picture}(0,0)(0,0)
   \put(-160,-10){$\phi_2$ (degree)}
   \put(-360,60){\rotatebox{90}{Br ($B\to \pi \rho$, $\pi\omega$) ($10^{-7}$)}}
  \end{picture}\end{center}
\caption{Branching ratios ($10^{-7}$) of $B^0\to \pi^0 \rho^0$  (solid line),
 $B^0\to \pi^0 \omega$ (dotted line),
as a function
of CKM angle $\phi_2$.}\label{00}
\end{figure}

  The averaged branching ratios of $B^0 \to \pi^0 \rho^0$ and $ \pi^0 \omega$
  are shown in Fig.\ref{00}.
  They also have a large dependence on $\phi_2$.
   The behavior of them is quite different, due to the different isospin
  of $\rho^0$ and $\omega$.
  But their branching ratios are rather small around $10^{-7}$.
  They may not be measured in the current running B factories,
  but may be possible in the future experiments, like LHC-B and NLC.
    The recent BABAR upper limit of the channel is   \cite{babar}
    \begin{equation}
    Br(B^0\to \pi^0\rho^0)   < 10.6 \times 10^{-6}.
    \end{equation}
   This is still consistent    with our predictions in standard model (SM).

\begin{figure}\begin{center}
\epsfig{file=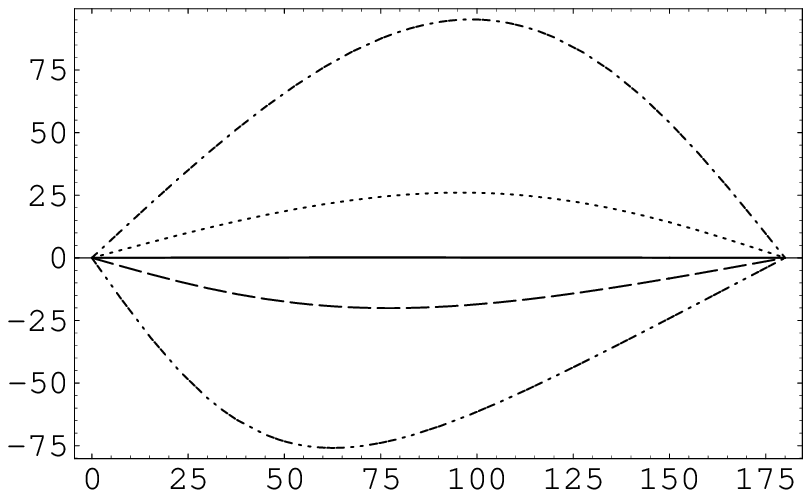,width=12cm}
\begin{picture}(0,0)(0,0)
   \put(-160,-10){$\phi_2$ (degree)}
   \put(-360,115){\rotatebox{90}{$A_{CP}^{Dir}$ (\%)}}
  \end{picture}\end{center}
\caption{Direct CP violation parameters of  $B^+\to \pi^+ \omega$ (solid
line),    $B^+\to \pi^+ \rho^0$
(dotted line), $B^+\to \rho^+ \pi^0$ (dashed line),
$B^0\to \rho^0 \pi^0$ (dash-dotted line), $B^0\to  \pi^0 \omega$
(dash-dotted-dotted line), as a function
of CKM angle $\phi_2$.}\label{cp00}
\end{figure}

Using eq.(\ref{m}-\ref{mb}), we can derive
the direct CP violating parameter as
\begin{eqnarray}
A_{CP}^{dir} &=& \frac{|{\cal M}|^2 -|\overline{\cal M}|^2}{
|{\cal M}|^2 +|\overline{\cal M}|^2}\nonumber\\ \label{dir}
&=& \frac{2 \sin \phi_2 \sin\delta}{1+2 z\cos \phi_2 \cos \delta +z^2}.
\end{eqnarray}
Unsurprisingly, it is a function of $\cos \phi_2$ and $\sin \phi_2$.
They are calculable in our PQCD approach.
The direct CP violation parameters as a function of $\phi_2$ are shown
in figure \ref{cp00}.
The direct CP violation parameter of $B^+\to \pi^+ \rho^0$ and
$B^0 \to \pi^0\rho^0$ are positive and very large.
While the direct CP violation parameter of $B^+\to \rho^+ \pi^0$ and
$B^0 \to \pi^0\omega$ are negative and very large.
The large strong phase required by the large direct CP asymmetry is
 from the non-factorizable and annihilation type diagrams, especially
the annihilation diagrams.
This is the different situation in Factorization
approach where the main contribution comes from BSS mechanism \cite{bss}
and the
annihilation diagram has been neglected \cite{akl2}.
The direct CP violation of $B^+\to \pi^+ \omega$ is rather small,
since the annihilation diagram contributions in this decay is almost
canceled out in eq.(\ref{omega}).
The preliminary measurement of CLEO shows a large CP asymmetry for this decay
\cite{cleocp}
\begin{equation}
A_{CP} (B^+\to \pi^+ \omega  )= 34 \pm 25 \% .
\end{equation}
Although the sign of CP is in agreement with our prediction, the central value
is too large.
If the result of central value remains in future experiments,
we may expect new physics contributions.

For the neutral $B^0$ decays, there is more complication from the
$B^0$ $\overline {B^0}$ mixing. The CP asymmetry is time dependent \cite{akl2}:
\begin{equation}
A_{CP} (t) \simeq A_{CP}^{dir} \cos (\Delta mt) + a_{\epsilon + \epsilon '}
\sin (\Delta m t),
\end{equation}
where $\Delta m$ is the mass difference of the two  mass eigenstates of
neutral $B$ meson.
The direct CP violation parameter $A_{CP}^{dir}$ has
already been defined in eq.(\ref{dir}). While the mixing-related CP violation
parameter is defined as
\begin{equation}
a_{\epsilon +\epsilon'}=\frac{ -2Im (\lambda_{CP})}
{1+|\lambda_{CP}|^2},
\end{equation}
where
\begin{equation}
\lambda_{CP} = \frac{ V_{tb}^*V_{td} \langle f |H_{eff}| \overline B^0\rangle}
{ V_{tb}V_{td}^* \langle f |H_{eff}| B^0\rangle}.
\end{equation}
Using equations (\ref{m},\ref{mb}), we can derive as
\begin{equation}
\lambda_{CP} = e^{2i\phi_2}\frac{ 1+ze^{i(\delta-\phi_2)} }{
 1+ze^{i(\delta+\phi_2)} }.
\end{equation}
$\lambda_{CP}$ and $a_{\epsilon +\epsilon'}$ are functions of CKM angle
$\phi_2$ only. Therefore, the CP asymmetry of $B\to \pi \rho$ and
$\pi\omega$ decays can measure the CKM angle $\phi_2$, even if for the
neutral B decays including the $B\bar B$ mixing effect.

\begin{figure}\begin{center}
\epsfig{file=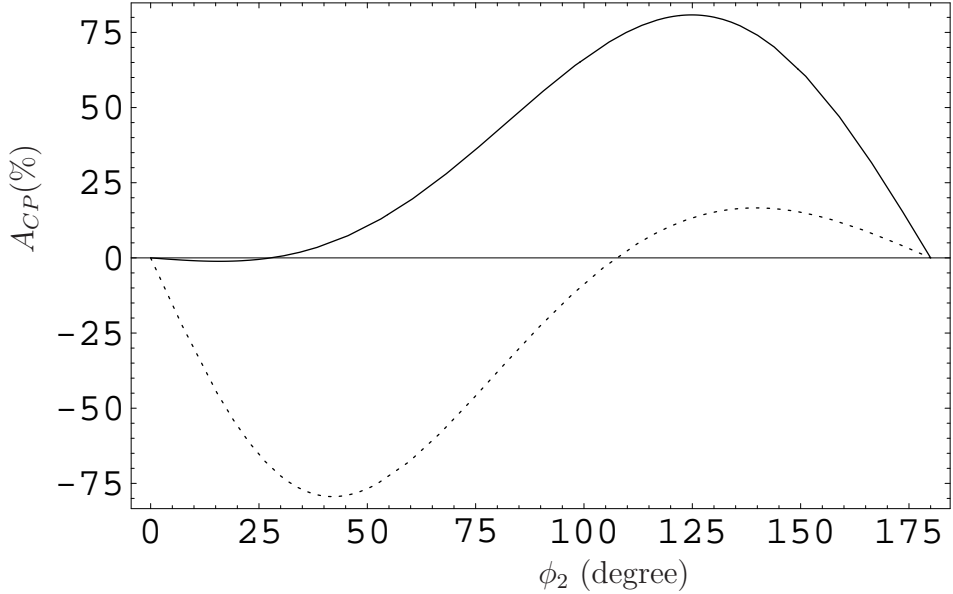,width=12cm}
\begin{picture}(0,0)(0,0)
   \put(-160,-10){$\phi_2$ (degree)}
   \put(-360,115){\rotatebox{90}{$A_{CP}$(\%)}}
  \end{picture}\end{center}
\caption{Total CP asymmetries of $B^0\to \pi^0 \rho^0$ (solid line),
and $B^0\to  \pi^0 \omega$ (dotted line),
as a function
of CKM angle $\phi_2$.}\label{cpome}
\end{figure}

If we integrate the time variable $t$, we will get the total CP asymmetry as
\begin{equation}
A_{CP} = \frac{1}{1+x^2} A_{CP}^{dir} + \frac{x}{1+x^2}a_{\epsilon +\epsilon'},
\end{equation}
with $x=\Delta m/\Gamma \simeq 0.723 $ for the $B^0 - \overline B^0$ mixing
in SM \cite{pdg}.
The total CP asymmetries of $B^0 \to \pi^0 \rho^0$, $\pi^0 \omega$ are shown
in Figure \ref{cpome}. Although the CP asymmetries are large, but it is still
difficult to measure for experiments, since their branching ratios are small
around $10^{-7}$.

\begin{figure}\begin{center}
\epsfig{file=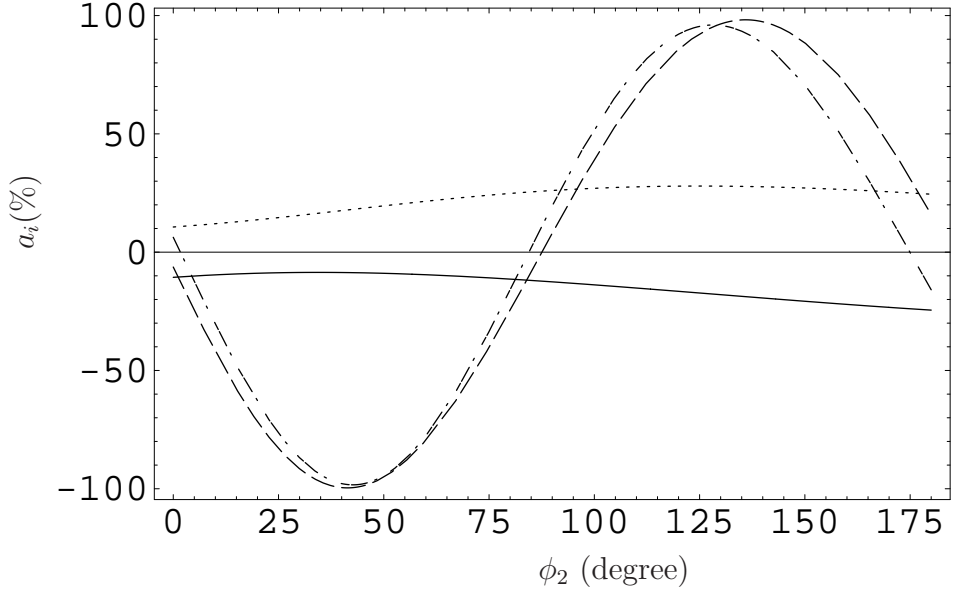,width=12cm}
\begin{picture}(0,0)(0,0)
   \put(-160,-10){$\phi_2$ (degree)}
   \put(-360,115){\rotatebox{90}{$a_{i}$(\%)}}
  \end{picture}\end{center}
\caption{CP violation parameters of $B^0/\bar B^0\to
\pi^\pm \rho^\mp$ decays: $a_{\epsilon '}$(solid line), $ a_{\bar \epsilon '}$
 (dotted line),
$a_{\epsilon +\epsilon '}$ (dashed line) and
$a_{\epsilon +\bar \epsilon '}$ (dash-dotted line)
as a function
of CKM angle $\phi_2$.}\label{cppm}
\end{figure}

The CP asymmetries of $B^0 /\bar B^0\to \pi^\pm \rho^\mp$ are very complicated.
 Here one studies the four
time-dependent decay widths for $B^0(t) \to \pi^+\rho^-$,
$\bar B^0(t) \to \pi^- \rho^+ $,
$B^0(t) \to \pi^- \rho^+ $ and $\bar B^0(t) \to  \pi^+\rho^-$
\cite{gr89,adkd91,PW95}.
These time-dependent widths can be expressed by four basic matrix elements
\begin{equation}
\begin{array}{ll}
g=\langle \pi^+\rho^-|H_{eff} |B^0\rangle  ,
& h=\langle \pi^+\rho^-|H_{eff}|\bar B^0\rangle ,\\
\bar g=\langle \pi^- \rho^+ |H_{eff} |\bar B^0\rangle  ,
& \bar h=\langle \pi^- \rho^+ |H_{eff}| B^0\rangle ,
\end{array}
\end{equation}
which determine the decay matrix elements of $B^0\to \pi^+\rho^-$
 \& $\pi^- \rho^+ $ and
of $\bar B^0 \to \pi^- \rho^+$  \& $\pi^+\rho^-$ at $t=0$.
The matrix elements $g$ and $\bar h$  are given in eq.(\ref{pm},\ref{pm2}).
The matrix elements $h$ and $\bar g$ are obtained from $\bar h$ and $g$ by
changing the signs of the weak phases contained in the products of the
CKM matrix elements. We also need to know  the CP-violating parameter
coming from the $B^0$ - $\bar{B}^0$ mixing. Defining:
\begin{eqnarray}
B_1 &=& p |B^0\rangle + q | \bar{B}^0 \rangle, \nonumber\\
B_2 &=& p |B^0\rangle - q | \bar{B}^0 \rangle,
\label{B12}
\end{eqnarray}
with $\vert p \vert^2 + \vert q \vert ^2=1$ and $q/p=\sqrt{H_{21}/H_{12}}$,
 with $H_{ij}= M_{ij} - i/2\Gamma_{ij}$ representing the
 $\vert \Delta B \vert =2$ and $\Delta Q=0$ Hamiltonian.
 For the decays of $B^0$ and $\bar{B}^0$, we use,
\begin{equation}
\frac{q}{p} = \frac{V_{tb}^* V_{td}}{V_{tb}V_{td}^*} = e ^{-2 i \phi_1}.
\label{qpdef}
\end{equation}
So, $|q/p|=1$, and this ratio has only a phase given by $-2 \phi_1$.
Then, the four time-dependent widths
are given by the following formulae (we follow the notation of \cite{PW95}):
\begin{eqnarray}
  \Gamma (B^0 (t) \to \pi^+\rho^-)
& =& e^{-\Gamma t} \frac{1}{2} ( |g|^2 + |h|^2 )
\left \{ 1+ a_{\epsilon '} \cos \Delta m t + a_{\epsilon +\epsilon '}
\sin \Delta m t \right \},\nonumber \\
 \Gamma (\bar B^0 (t) \to \pi^- \rho^+ )& =& e^{-\Gamma t} \frac{1}{2} ( |\bar g|^2
+ |\bar h|^2 ) \left \{ 1- a_{\bar \epsilon '} \cos \Delta m t
- a_{\epsilon +\bar \epsilon '}
\sin \Delta m t \right \},\nonumber \\
 \Gamma ( B^0 (t) \to \pi^- \rho^+ ) &= &e^{-\Gamma t} \frac{1}{2} ( |\bar g|^2
+ |\bar h|^2 )\left \{ 1+ a_{\bar \epsilon '}
 \cos \Delta m t + a_{\epsilon +\bar \epsilon '}
\sin \Delta m t \right \},\nonumber \\
 \Gamma (\bar B^0 (t) \to \pi^+\rho^-) &=& e^{-\Gamma t} \frac{1}{2} ( | g|^2
+ | h|^2 )
\left \{ 1- a_{\epsilon '} \cos \Delta m t - a_{\epsilon +\epsilon '}
\sin \Delta m t \right \},  \label{rate}
\end{eqnarray}
where
\begin{equation}
  \label{aepsilon}
  \begin{array}{ll}
a_{\epsilon '} = \displaystyle \frac{ |g|^2 -|h|^2}{ |g|^2 +|h|^2}, &
a_{\epsilon +\epsilon '} = \displaystyle
\frac{-2Im \left( \frac{q}{p}\frac{h}{g}\right)}
{1+|h/g|^2},\\
a_{\bar \epsilon '} = \displaystyle
\frac{ |\bar h|^2 -|\bar g|^2}{ |\bar h|^2 +|\bar g|^2},
 &
a_{\epsilon +\bar \epsilon '} = \displaystyle
\frac{-2Im \left( \frac{q}{p}\frac{\bar g}
{\bar h}\right)}
{1+|\bar g/\bar h |^2}.
   \end{array}
\end{equation}
We calculate the above four CP violation parameters related to
$B^0 /\bar B^0 \to \pi^\pm \rho^\mp$ decays in PQCD.
The results are shown in Fig.\ref{cppm} as a function of $\phi_2$.
 Comparing the results with the factorization approach \cite{akl2},
we found that
our predicted size of $a_{\epsilon '}$ and $a_{\bar \epsilon '}$ are smaller
while $a_{\epsilon +\epsilon'}$  and $a_{\epsilon +\bar \epsilon'}$
are larger.
By measuring the time-dependent spectrum of the decay rates of $B^0$ and
$\bar B^0$, one can find the coefficients of the two functions $\cos\Delta mt$
and $\sin \Delta m t$ in eq.(\ref{rate})
and extract the quantities $a_{\epsilon '}$,
$a_{\epsilon +\epsilon '}$,  $a_{\bar \epsilon '}$, and
$a_{\epsilon +\bar \epsilon '}$. Using these experimental results, we can tell
the size of CKM angle $\phi_2$ from Fig.\ref{cppm}.

\section{Summary}

We calculate the $B^0 \to \pi^+\rho^-$,
$B^0 \to \rho^+\pi^-$, $B^+ \to \rho^+\pi^0$,
$B^+ \to \pi^+\rho^0$, $B^0 \to \pi^0\rho^0$, $B^+ \to \pi^+\omega$
and $B^0 \to \pi^0\omega$ decays, together with their
charge conjugate modes, in a  perturbative QCD approach.
We calculate the $B\to \pi$ and $B\to \rho$ form factors, which are in
agreement with the QCD sum rule calculations.
In addition to the usual factorization contributions, we also
calculate the non-factorizable and annihilation diagrams.
Although they are sub-leading contributions in the branching ratios of
these decays, they are not negligible.
Furthermore these diagrams provide the necessary strong phases required
by the direct CP asymmetry measurement.

Our calculation gives the right branching ratios, which agrees well with the
CLEO and BABAR measurements. We also predict large direct CP asymmetries in
$B^+ \to \rho^+\pi^0$ and $B^+ \to \pi^+\rho^0$ decays.
Including the $B\bar B$ mixing effect, the CP asymmetries of
$B^0 \to \pi^0 \omega$ and $B^0 \to \pi^0\rho^0$ are very large, but their
branching ratios are small in SM.
The CP asymmetry parameters of  $B^0 \to \pi^+\rho^-$,
$B^0 \to \rho^+\pi^-$ decays require the time dependent measurement of
branching ratios.

\section*{Acknowledgments}

We thank our PQCD group members: Y.Y. Keum, E. Kou, T. Kurimoto, H.N. Li,
T. Morozumi,
A.I. Sanda, N. Sinha, R. Sinha, K. Ukai and T. Yoshikawa
 for helpful discussions.
This work was supported by the Grant-in-Aid for Scientific Research
on Priority Areas (Physics of CP violation).
We also thank JSPS for support.

\begin{appendix}

\section{Related functions defined in the text}

We show here
 the function $h_i$'s, coming from the
Fourier transform of $H^{(0)}$,
\begin{eqnarray}
h_e(x_1,x_2,b_1,b_2)&=&
 K_{0}\left(\sqrt{x_1x_2} m_B b_1\right)
 \left[\theta(b_1-b_2)K_0\left(\sqrt{x_2} m_B
b_1\right)I_0\left(\sqrt{x_2} m_B b_2\right)\right.
\nonumber \\
& &\;\;\;\;\left.
+\theta(b_2-b_1)K_0\left(\sqrt{x_2}  m_B b_2\right)
I_0\left(\sqrt{x_2}  m_B b_1\right)\right]  S_t(x_2)
\label{ha}\\
h_{d}(x_1,x_2,x_3,b_1,b_2)&=&
\alpha_{s}(t_{d})K_{0}\left(-i\sqrt{x_2 x_3} m_B b_2\right)
\nonumber \\
& &\times \left[\theta(b_1-b_2)K_0\left(\sqrt{x_1 x_2} m_B
b_1\right)I_0\left(\sqrt{x_1 x_2} m_B b_2\right)\right.
\nonumber \\
& &\left.
+\theta(b_2-b_1)K_0\left(\sqrt{x_1 x_2}  m_B b_2\right)
I_0\left(\sqrt{x_1 x_2}  m_B b_1\right)\right]
\label{hd}\\
h_{f}^1(x_1,x_2,x_3,b_1,b_2)&=&
K_{0}\left(-i\sqrt{x_2 x_3} m_B b_1\right)
\alpha_{s}(t_{f}^1) \nonumber \\
& &\times \left[\theta(b_1-b_2)K_0\left(-i\sqrt{x_2 x_3} m_B
b_1\right)J_0\left(\sqrt{x_2 x_3} m_B b_2\right)\right.
\nonumber \\
& &\left.
+\theta(b_2-b_1)K_0\left(-i\sqrt{x_2 x_3}  m_B b_2\right)
J_0\left(\sqrt{x_2 x_3}  m_B b_1\right)\right]\label{hf}
\\
h_{f}^2(x_1,x_2,x_3,b_1,b_2)&=&
K_{0}\left(\sqrt{x_2+x_3-x_2 x_3} m_B b_1\right)\alpha_{s}(t_{f}^2)
\nonumber \\
& &\times \left[\theta(b_1-b_2)K_0\left(-i\sqrt{x_2 x_3} m_B
b_1\right)J_0\left(\sqrt{x_2 x_3} m_B b_2\right)\right.
\nonumber \\
& &\left.
+\theta(b_2-b_1)K_0\left(-i\sqrt{x_2 x_3}  m_B b_2\right)
J_0\left(\sqrt{x_2 x_3}  m_B b_1\right)\right]\label{he}
\\
h_{a}(x_1,x_2,b_1,b_2)&=&
K_0 \left(-i\sqrt{x_1 x_2} m_B b_2\right)   S_t(x_1)
\nonumber \\
& &\times
\left[\theta(b_1-b_2)K_0\left(-i\sqrt{x_1 } m_B
b_1\right)J_0\left(\sqrt{x_1 } m_B b_2\right)\right.
\nonumber \\
& &\left.~~~~
+\theta(b_2-b_1)K_0\left(-i\sqrt{x_1 }  m_B b_2\right)
J_0\left(\sqrt{x_1 }  m_B b_1\right)\right],
\end{eqnarray}
where $J_0$ is the Bessel function and  $K_0$, $I_0$ are
modified Bessel functions $K_0 (-i x) = -(\pi/2) Y_0 (x) + i (\pi/2)
J_0 (x)$.    The threshold resummation form factor $S_t(x_i)$ is
adopted from ref.\cite{kurimoto}
\begin{equation}
S_t(x)=\frac{2^{1+2c} \Gamma (3/2+c)}{\sqrt{\pi} \Gamma(1+c)}
[x(1-x)]^c,
\end{equation}
where the parameter $c=0.3$. This function is normalized to unity.

The Sudakov factors used in the text are defined as
\begin{eqnarray}
S_{ab}(t) &=& s\left(x_1 m_B/\sqrt{2}, b_1\right)
+s\left(x_2 m_B/\sqrt{2}, b_2\right)
+s\left((1-x_2) m_B/\sqrt{2}, b_2\right) \nonumber \\
& &-\frac{1}{\beta_1}\left[\ln\frac{\ln(t/\Lambda)}{-\ln(b_1\Lambda)}
+\ln\frac{\ln(t/\Lambda)}{-\ln(b_2\Lambda)}\right],
\label{wp}\\
S_{cd}(t) &=& s\left(x_1 m_B/\sqrt{2}, b_1\right)
 +s\left(x_2 m_B/\sqrt{2}, b_2\right)
+s\left((1-x_2) m_B/\sqrt{2}, b_2\right) \nonumber \\
&& +s\left(x_3 m_B/\sqrt{2}, b_1\right)
+s\left((1-x_3) m_B/\sqrt{2}, b_1\right) \nonumber \\
& &-\frac{1}{\beta_1}\left[2 \ln\frac{\ln(t/\Lambda)}{-\ln(b_1\Lambda)}
+\ln\frac{\ln(t/\Lambda)}{-\ln(b_2\Lambda)}\right],
\label{Sc}\\
S_{ef}(t) &=& s\left(x_1 m_B/\sqrt{2}, b_1\right)
 +s\left(x_2 m_B/\sqrt{2}, b_2\right)
+s\left((1-x_2) m_B/\sqrt{2}, b_2\right) \nonumber \\
&& +s\left(x_3 m_B/\sqrt{2}, b_2\right)
+s\left((1-x_3) m_B/\sqrt{2}, b_2\right) \nonumber \\
& &-\frac{1}{\beta_1}\left[\ln\frac{\ln(t/\Lambda)}{-\ln(b_1\Lambda)}
+2\ln\frac{\ln(t/\Lambda)}{-\ln(b_2\Lambda)}\right],
\label{Se}\\
S_{gh}(t) &=& s\left(x_2 m_B/\sqrt{2}, b_1\right)
 +s\left(x_3 m_B/\sqrt{2}, b_2\right)
+s\left((1-x_2) m_B/\sqrt{2}, b_1\right) \nonumber \\
&+&
s\left((1-x_3) m_B/\sqrt{2}, b_2\right)
-\frac{1}{\beta_1}\left[\ln\frac{\ln(t/\Lambda)}{-\ln(b_1\Lambda)}
+\ln\frac{\ln(t/\Lambda)}{-\ln(b_2\Lambda)}\right],
\label{ww}
\end{eqnarray}
where the function $s(q,b)$ are defined in the Appendix A of ref.\cite{luy}.
The scale $t_i$'s in the above equations are chosen as
\begin{eqnarray}
t_{e}^1 &=& {\rm max}(\sqrt{x_2} m_B,1/b_1,1/b_2)\;,\\
t_{e}^2 &=& {\rm max}(\sqrt{x_1}m_B,1/b_1,1/b_2)\;,\\
t_{d} &=& {\rm max}(\sqrt{x_1x_2}m_B,
\sqrt{x_2x_3} m_B,1/b_1,1/b_2)\;,\\
t_{f}^1 &=& {\rm max}(\sqrt{x_2x_3}m_B,
1/b_1,1/b_2)\;,\\
t_{f}^2 &=& {\rm max}(\sqrt{x_2x_3}m_B,\sqrt{x_2+x_3-x_2 x_3} m_B,
1/b_1,1/b_2)\;.
\end{eqnarray}

\end{appendix}


\end{document}